\documentclass[11pt]{article}
\usepackage{amsfonts}
\usepackage{graphicx}
\usepackage{amssymb}
\usepackage{authblk}
\usepackage{amsmath}
\usepackage{mathtools}
\usepackage{mathrsfs}
\usepackage{multirow}
\usepackage{algorithm}
\usepackage{algpseudocode}
\usepackage{placeins}

\usepackage{bm}

\usepackage{url}
\usepackage{scrextend}
\usepackage{longtable}
\usepackage{caption}
\usepackage{multirow}
\usepackage[colorinlistoftodos]{todonotes}
\usepackage{algpseudocode}
\usepackage{dcolumn}
\usepackage{url}

\def\bSig\mathbf{\Sigma}

\algdef{SE}[SUBALG]{Indent}{EndIndent}{}{\algorithmicend\ }%
\algtext*{Indent}
\algtext*{EndIndent}

\makeatletter     
\newcommand{\distas}[1]{\mathbin{\overset{#1}{\kern\z@\sim}}}%
\newsavebox{\mybox}\newsavebox{\mysim}
\newcommand{\distras}[1]{%
  \savebox{\mybox}{\hbox{\kern3pt$\scriptstyle#1$\kern3pt}}%
  \savebox{\mysim}{\hbox{$\sim$}}%
  \mathbin{\overset{#1}{\kern\z@\resizebox{\wd\mybox}{\ht\mysim}{$\sim$}}}%
}
\makeatother

\DeclareMathOperator*{\W}{\mathbf{W}}

\DeclareMathOperator*{\Wxl}{\mathbf{W}^T\mathbf{x}_{\ell}}
\DeclareMathOperator*{\yl}{\mathbf{y}_{\ell}}

\DeclareMathOperator*{\prodid}{\prod_{i=1}^d}

\DeclareMathOperator*{\sumln}{\sum_{\ell=1}^n}

\DeclareMathOperator*{\sumid}{\sum_{i=1}^d}

\DeclareMathOperator*{\bvmui}{\mathbf{\underset{\sim}{\mu_i}}}

\DeclareMathOperator*{\omegai}{\omega_i^2}
\DeclareMathOperator*{\omegad}{\omega_d^2}

\DeclareMathOperator*{\omegao}{\omega_1^2}

\DeclareMathOperator*{\vomega}{\boldsymbol{\omega^2}}

\def \vTheta{\mbox{\boldmath $\Theta$ \unboldmath}\!\!}

\def \vW{\mbox{\boldmath $W$ \unboldmath}\!\!}

\def \vSigma{\mbox{\boldmath $\Sigma$ \unboldmath}\!\!}
\def \vTheta{\mbox{\boldmath $\Theta$ \unboldmath}\!\!}

\def \vomega{\mbox{\boldmath $\omega$ \unboldmath}\!\!}

\def \vTheta{\mbox{\boldmath $\Theta$ \unboldmath}\!\!}
\def \vomega{\mbox{\boldmath $\omega$ \unboldmath}\!\!}
\def \vmu{\mbox{\boldmath $\mu$ \unboldmath}\!\!}

\def \vY{\mbox{\boldmath $Y$ \unboldmath}\!\!}

\def \vtheta{\mbox{\boldmath $\theta$ \unboldmath}\!\!}

\usepackage[default]{jasa_harvard}    
\usepackage{JASA_manu}

\newcommand*\samethanks[1][\value{footnote}]{\footnotemark[#1]}

\begin{document}

\title{A Bayesian Spatial Model for Imaging Genetics}

\author[1]{Yin Song\thanks{ The authors wish it to be known that the first two authors
should be regarded as joint First Authors.}} 
\author[2]{Shufei Ge\samethanks} 
\author[2]{Jiguo Cao}
\author[2]{Liangliang Wang}
\author[1]{Farouk S. Nathoo\thanks{Corresponding Author: nathoo@uvic.ca.}} 

\affil[1]{Department of Mathematics and Statistics, University of Victoria}
\affil[2]{Department of Statistics and Actuarial Science, Simon Fraser University}

\maketitle

\begin{center}
\textbf{Abstract}
\end{center}We develop a Bayesian bivariate spatial model for multivariate regression analysis applicable to studies examining the influence of genetic variation on brain structure. Our model is motivated by an imaging genetics study of the Alzheimer's Disease Neuroimaging Initiative (ADNI), where the objective is to examine the association between images of volumetric and cortical thickness values summarizing the structure of the brain as measured by magnetic resonance imaging (MRI) and a set of 486 SNPs from 33 Alzheimer's Disease (AD) candidate genes obtained from 632 subjects. A bivariate spatial process model is developed to accommodate the correlation structures typically seen in structural brain imaging data. First, we allow for spatial correlation on a graph structure in the imaging phenotypes obtained from a neighbourhood matrix for measures on the same hemisphere of the brain. Second, we allow for correlation in the same measures obtained from different hemispheres (left/right) of the brain. We develop a mean-field variational Bayes algorithm and a Gibbs sampling algorithm to fit the model. We also incorporate Bayesian false discovery rate (FDR) procedures to select SNPs. We implement the methodology in a new a release of the R package \emph{bgsmtr}. We show that the new spatial model demonstrates superior performance over a standard model in our application. Data used in the preparation of this article were obtained from the Alzheimer's Disease Neuroimaging Initiative (ADNI) database (adni.loni.usc.edu).


\newpage

\section{Introduction}
\label{s:intro}
We consider multivariate multiple regression modeling within the context of imaging genetics where interest lies in uncovering the associations between genetic variations and neuroimaging measures as quantitative traits (QTs). This problem has received a great deal of attention recently and is challenging because it combines the analysis of neuroimaging data with genetic data (see e.g., Vounou et al., 2010; Stein et al., 2010; Silver et al., 2010; Inkster et al., 2010; Hibar et al., 2011; Ge et al., 2012; Thompson et al., 2013; Stingo et al., 2013;  Zhu et al., 2014; Hibar et al., 2015; Huang et al., 2015; Huang et al., 2017; Lu et al., 2017). Recent reviews of statistical issues in this area are discussed in Liu and Calhoun (2014) and Nathoo et al. (2018). 

The neuroimaging measures can serve as endophenotypes for neurological disorders such as Alzheimer's disease (AD). As described in Szefer et al. (2017), the estimated heritability of late-onset AD is 60 - 80 percent (Gatz et al., 2006). The largest susceptibility allele is the $\epsilon$4 allele of the Apolipoprotein E gene (APOE; Corder et al. 1993), which may play a role in 20 to 25 percent of AD cases. The remaining heritability of AD may be explained by many additional genetic variants and these may have a small effect.

In our work, we consider the setting where interest lies in assessing the association between a moderate number of brain imaging phentoypes (e.g., 111 ROIs in Vounou et al., 2010; 12 ROIs in Wang et al., 2012; 93 ROIs in Zhu et al., 2014; 56 ROIs in Greenlaw et al., 2017) and with the number of SNPs ranging from between a few hundred to a few thousand. Within this setting a multivariate model with regression matrix jointly characterizing the associations between all ROIs and genetic markers is feasible.

Greenlaw et al. (2017) propose a Bayesian group sparse multi-task regression model where the primary focus is the use of a shrinkage prior based on a product of multivariate Laplace kernels developed following the ideas of Park and Casella (2008) and Kyung et al. (2010). The specific prior developed is motivated by the penalized multi-task regression estimator proposed by Wang et al. (2012). This development is an effort to move from point estimation to Bayesian credible intervals and fully Bayesian inference.

While these authors demonstrate the advantage of characterizing posterior uncertainty in their imaging genetics application to the ADNI study, their model makes a simplifying assumption for the covariance matrix of the imaging phenotypes, where the first level of the model assumes:
\begin{equation} \yl |\mathbf{W},\sigma^2  \distas{ind} MVN_c (\Wxl \: , \: {\sigma^2I_c}) \hspace{8pt} \ell=1, \dots, n, \end{equation}
where $\yl = (\mathbf{y}_{\ell 1}, \dots , \mathbf{y}_{ \ell c})^T$ denotes the vector of imaging phenotypes for subject $\ell$ and $c$ is the dimension of the imaging phenotype, where $\ell = 1,\dots, n$; $\mathbf{A}^T$ denotes transpose of matrix $\mathbf{A}$, $\mathbf{W}$ is the regression matrix; $\mathbf{x}_{\ell} = (\mathbf{x}_{\ell 1}, \dots , \mathbf{x}_{\ell d})^T$, where $\mathbf{x}_{\ell}$ denotes the vector of genetic markers for subject $\ell$ and $d$ is the number of such markers. The assumed covariance structure ignores spatial correlation as well as bilateral correlation across brain hemispheres. 

We develop a new model that allows for this type of correlation by adopting a proper bivariate conditional autoregressive process (BCAR; see, e.g., Gelfand and Vounatsou, 2003; Jin et al., 2005) for the errors in the regression model. While spatial models for functional magnetic resonance imaging (fMRI) and other neuroimaging modalities have been developed to a large extent (see, e.g., Penny et al., 2005; Bowman, 2005; Bowman et al., 2008; Derado et al., 2013; Teng et al., 2018a; Teng et al., 2018b), to our knowledge there has been very little development of explicitly spatial models for imaging genetics. One exception is the mixture model developed by Stingo et al. (2013) where an Ising prior, a binary Markov random field, is used for Bayesian variable selection. Our model is rather different in both its aims and structure as it is based on a continuous bivariate Markov random field that is specified at the first level of the model for the imaging phenotype directly. 

In Figure 1 we show several summaries of the data from our motivating application demonstrating the need to account for correlation across brain hemispheres. For example, the sample correlation between the volume of the right cerebral cortex and the volume of the left cerebral cortex is 0.90, and across all 28 pairs of measurements, the median left/right correlation between corresponding phenotypes is approximately 0.8. The bivariate CAR structure allows us to account for this between-hemisphere correlation while also allowing us to account for within-hemisphere correlation using a graph structure based on a neighbourhood matrix.

\begin{figure}[h]
\centering
\caption{The correlation plot for left and right brain measures showing the bilateral correlation between 28 brain measurements across brain hemispheres for 632 subjects. Panel (a) presents the boxplot of all of the pairwise bilateral correlations for the 28 brain measurements across left and right hemispheres. Panel (b) presents a scatter plot showing mean thickness of left/right frontal with correlation 0.91. Panel (c) presents a scatter plot showing the volume of left/right cerebral cortex with correlation 0.90. Panel (d) presents a scatter plot between volume of left/right cerebral white matter with correlation 0.90.}	
\includegraphics[scale=0.35]{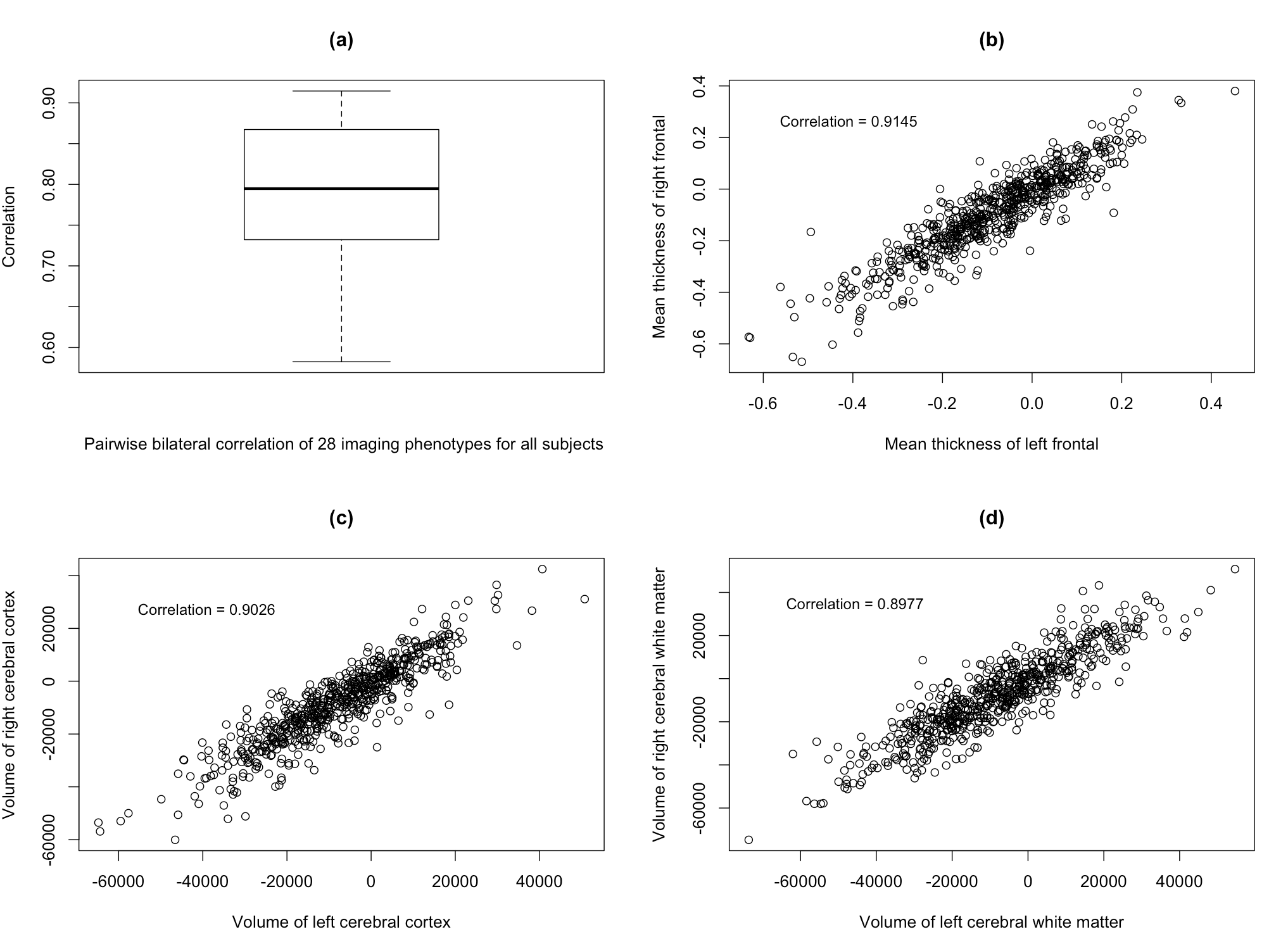}
\label{snpregion}
\end{figure}

Typically, models incorporating multivariate CAR specifications are used for modelling observations (in the case of a proper CAR model) or spatially-varying parameters when multiple observations or parameters appear at each spatial site. For our application the use of this process is non-standard in the sense that we do not model multiple observations at each site, but rather, we pair corresponding observations on opposite hemispheres of the brain and use the bivariate spatial process to model a combination of the bilateral correlation across the left and right brain hemispheres as well as potential correlation within each hemisphere. As a matter of fact for the MRI data considered in our application the bilateral correlation is a very strong signal in the observed data and so it is important to account for it. 

For the bivariate spatial model we use a separable BCAR process as it is reasonable in our application to assume (as it might be in other neuroimaging studies) that the spatial covariance on the two hemispheres of the brain is similar. Non-separable multivariate spatial models (see, e.g., Gelfand and Banerjee, 2010; MacNab 2016) could be adopted for more flexibility allowing the spatial structure on the two hemispheres to be different; however, we do not expect that this additional flexibility would be useful in the current context. This spatial process is combined with a group Lasso prior for the regression coefficients, where each group corresponds to a single row of $\mathbf{W}$. Each row in this case represents the associations between a given SNP and all of the phenotypes. 

To compute the posterior distribution we develop two algorithms, both of which are implemented in our R package \emph{bgsmtr} for imaging genetics regression modelling. The package is available for download on the Comprehensive R Archive Network (CRAN). The first algorithm is a Gibbs sampling algorithm and the second is a faster mean-field variational Bayes (VB) approximation to the posterior distribution (see e.g., Ormerod and Wand, 2010). Within the context of hierarchical models for spatial data, mean-field VB inference has been considered by Ren et al. (2011) who make comparisons with inference from MCMC within the context of spatial process models. In addition to the computation of the posterior distribution, the \emph{bgsmtr} package now incorporates Bayesian FDR procedures (Morris et al., 2008) for SNP selection. This can be used alongside or as an alternative to SNP selection based on credible intervals.

The overall contribution of our work is four-fold. First, we develop an explicitly spatial model for imaging genetics based on the BCAR process. Second, we develop both an MCMC algorithm and a mean-field VB algorithm for approximating the posterior distribution. Third, we incorporate Bayesian FDR procedures for SNP selection within the new spatial model. Fourth, our new developments are implemented in the latest version of the \emph{bgsmtr} R package.

The remainder of this paper is structured as follows. In Section 2, we present our new spatial model for imaging genetics. Computation of the posterior distribution and SNP selection is discussed in Section 3. Section 4 presents a simulation study evaluating the performance of the spatial model relative to a non-spatial model and inference based on MCMC relative to that from VB. In Section 5 we apply our new model to our motivating application examining data from the ADNI-1 study, examining 56 structural brain imaging phenotypes, 486 SNPs from 33 genes, and 632 subjects. The paper concludes with a discussion in Section 6.

\section{Bayesian Spatial Regression Model}
Let $\yl = (\mathbf{y}_{\ell, 1}, \dots , \mathbf{y}_{ \ell, c})^T$ and $\mathbf{x}_{\ell} = (\mathbf{x}_{\ell ,1}, \dots , \mathbf{x}_{\ell ,d})^T$ denote the imaging measures at $c$ ROIs and the genetic data respectively for subject $\ell$, $\ell = 1,\dots, n$, where $\mathbf{x}_{\ell, j} \in \{0,1,2\}$ represents the number of minor alleles of the $j_{th}$ SNP for subject $\ell$. The regression model takes the form $E(\mathbf{y_{\ell}}) = \vW^T\mathbf{x}_\ell, \;\ell = 1,\dots, n$, where $\vW$ has dimensions $d \times c$ and $W_{i,j}$ represents the association between the $i_{th}$ SNP and the $j_{th}$ imaging phenotype. 

Our model is developed for settings where the imaging data are symmetric with the same measures collected on each hemisphere of the brain. This is true when the neuroimaging data are considered at the voxel level and it is also the case for the study considered here where we analyze MRI data from the ADNI-1 database preprocessed using the FreeSurfer V4 software (Fischl, 2012). We conduct automated parcellation to define volumetric and cortical thickness values from the $28$ ROIs considered in Shen et al. (2010), Szefer et al. (2017), and Greenlaw et al. (2017) on each hemisphere leading to $c=56$ brain measures in total. 

As described in Szefer et al. (2017), potential confounders in the analysis are population stratification and APOE genotype. Since true population structure is not observed, a set of principal coordinates from multidimensional scaling are used to derive proxy variables for population stratification in the data. We also adjust for APOE genotype, since it can account for the population stratification in the data, over and above the principal components or principal coordinates (Lucotte et al. 1997).

The response imaging measures at each brain ROI are first adjusted for the ten principal coordinates, as well as for dummy variables representing APOE genotype, using weighted ordinary least squares regression. The residuals from each regression are then used as the adjusted neuroimaging phenotypes (Szefer et al., 2017).

Let $\boldsymbol{y}_{\ell, i} = (y_{l,i}^{(L)}, y_{l,i}^{(R)})'$ be the brain summary measures obtained at the $i_{th}$ ROI in the left hemisphere (L) and the right hemisphere (R). Then
 $\boldsymbol{y}_{\ell } =(\boldsymbol{y}_{\ell, 1}',\dots, \boldsymbol{y}_{\ell ,c/2}' )'$ is the imaging data ordered so that left-right imaging phenotype pairs are adjacent in the response vector. There are thus $c/2$ ROIs on each hemisphere and we let $\boldsymbol{A}$ denote a $c/2 \times c/2$ symmetric neighborhood matrix which in the simplest case can have binary elements, where $A_{i,j} = 1$ indicates that ROI $i$ and $j$ are neighbors $i \ne j$, or more generally $A_{i,j} \ge 0$ and $A_{i,i} = 0$, $i=1,\dots,c/2$. A user input is thus a neighborhood matrix $\boldsymbol{A}$ or in the absence of user input our software implementation takes $A_{i,j}$ to be the average of the absolute value of the sample correlation between phenotype/ROI $i$ and phenotype/ROI $j$, where the average is taken over left/right hemisphere. The regression model then takes the form
 \vspace{-1em}
 \begin{equation}
\boldsymbol{y}_{\ell} = \vW^T \boldsymbol{x}_{\ell} + \boldsymbol{\epsilon}_{\ell} 
\end{equation} 
 and the model for the errors $\boldsymbol{\epsilon}_{\ell}$ is a mean-zero multivariate normal distribution of dimension $c$, which can be specified through a set of $c/2$ compatible bivariate conditional distributions for $\boldsymbol{\epsilon}_{l,i} = (\epsilon_{l,i}^{(L)}, \epsilon_{l,i}^{(R)})'$, specified as follows:
\vspace{-1em}
    \begin{equation*}
	\boldsymbol{\epsilon}_{l,i} |\boldsymbol{\epsilon}_{l(-i)}, \rho,\boldsymbol{\Sigma} \sim \text{BVN}\big( \frac{\rho}{{A}_{i.}} \sum_{j=1}^{c/2}A_{i,j}\boldsymbol{\epsilon_{l,j}}, \frac{1}{{A}_{i.}}\boldsymbol{\Sigma} \big) 
	\end{equation*} 	
where, $\boldsymbol{\epsilon_{l(-i)}}$ denotes the rest after removing $\boldsymbol{\epsilon}_{l,i}$ from $\boldsymbol{\epsilon}_{l}$, $A_{i.}=\sum_{j=1}^{\frac{c}{2}}A_{i,j}$,  $\rho \in [0, 1)$ characterizes spatial dependence with $\rho=0$ corresponding to independence across all ROI pairs and $\boldsymbol{\Sigma}$ is a $2 \times 2$ matrix where $\kappa = \Sigma_{12} / \sqrt{\Sigma_{11}\Sigma_{22}} \in (-1, 1)$ quantifies within pair dependence, and with $\kappa = 0$ corresponding to independence within ROI pairs. 

As far as we are aware, this spatial model for neuroimaging data is one of the first to explicitly model dependence across brain hemispheres in addition to accounting for local dependence. Often, this left/right bilateral dependence is ignored with neuroimaging data. In our data it is a very clear and strong signal as is evident in Figure 1. As the parameter $\Sigma$ is free in the model it can be informed by the data. Therefore, we expect that the posterior will reflect some degree of between hemisphere correlation when it is present to a sufficient degree, and will remain at roughly a diagonal form when it is not, in diseases that have major differences across hemispheres, for example. The prior for $\Sigma$ is chosen so that it is centered on a diagonal matrix.

Under this new specification the first level of the regression model takes the following form:
\begin{equation}
\label{CAR_MVN}
\boldsymbol{y}_{\ell} | \boldsymbol{W}, \boldsymbol{\Sigma} \stackrel{ind}{\sim} \text{MVN}_{c}(\boldsymbol{W}^{T}\boldsymbol{x}_{\ell}, { {(\boldsymbol{D_{A}} - \rho \boldsymbol{A})^{-1} \otimes  \boldsymbol{\Sigma}}) }, \, l=1,\dots,n,
\end{equation}
\vspace{0.5em}
where $\text{MVN}_{c}$ denotes a $c$-dimensional multivariate normal distribution, $\otimes$ is the kronecker product, $\boldsymbol{D_{A}} = \text{diag}\{A_{i.}, i = 1, ..., c/2\}$ and as before $A_{i.}=\sum_{j=1}^{\frac{c}{2}}A_{i,j}.$ For the regression coefficients, we let
$\tilde {W}_{i,j^{*}}=(W_{i,j},W_{i,{j+1}})$,~~$j=2j^{*}-1$,~~$j^{*}=1,...,\frac{c}{2}$, and we adopt a shrinkage prior based on a bivariate Gaussian scale mixture
\begin{equation*}
\tilde{W}_{i,j^{*}}| \omega_{i}^{2}, \Sigma \stackrel{ind}{\sim}\text{BVN}(\boldsymbol{0},\omega_{i}^2\Sigma), 
\end{equation*}
\begin{equation*}
\omega_{i}^{2}|\lambda^{2} \stackrel{iid}{\sim} \text{Gamma}
(\frac{c+1}{2},\lambda^{2}/2), \,\, \Sigma \sim \text{Inv-Wishart}(v,\boldsymbol{S}),
\end{equation*}
where $\rho$ and $\lambda^{2}$ are tuning parameters controlling spatial dependence and regression sparsity respectively. Tuning of the model is discussed in Section 3.3 after our discussion of computational algorithms for fitting the model. The remaining hyperparameters $v$ and $\boldsymbol{S}$ are set at $v=2$ and $\boldsymbol{S} = \boldsymbol{I}$ to yield a prior that is somewhat vague, and they can be varied as part of a sensitivity analysis. 

\section{Computation and SNP Selection}

\subsection{Bayesian Computation}

Posterior computation can be implemented using Gibbs sampling. The update steps for this algorithm are listed in Algorithm 1 and their derivations are given in the Supplementary Material (Web Appendix B). As a faster alternative approach to computing the posterior distribution, we also develop a mean-field VB algorithm. The approximation $q(\vtheta)$ to the posterior distribution $P(\vtheta|\mathbf{Y})$  is based on constructing and optimizing a lower bound on the marginal likelihood $P(\mathbf{Y})$. Assuming that $q(\vtheta)$ has the same support as $P(\vtheta|\mathbf{Y})$, the log-marginal likelihood can be written as $\text{log} P(\mathbf{Y})$
\vspace{-1em}
 \begin{align*}
    &= \int q(\vtheta)\log\{\frac{P(\mathbf{Y},\vtheta)}{q(\vtheta)}\}d\vtheta + \int q(\vtheta)\log\{\frac{q(\vtheta)}{P(\vtheta|\mathbf{Y})}\}d\vtheta\\
&= E_{q}[\log\{\frac{P(\vtheta, \mathbf{Y})}{q(\vtheta)}\}] + E_{q}[\log\{\frac{q(\vtheta)}{P(\vtheta|\mathbf{Y})}\}]  \\
&= \mathfrak{F}(q,\mathbf{Y}) + KL(q||p) \ge \mathfrak{F}(q,\mathbf{Y}).
\end{align*}

Here, $KL(q||p)$ denotes the Kullback-Leibler divergence from $q$ to $p$ and the final inequality is true since $KL(q||p) \ge 0$. The approximation to $P(\vtheta|\mathbf{Y})$ by $q(\vtheta)$ is obtained by restricting $q(\vtheta)$ to a manageable class of distributions and maximizing the lower bound $\mathfrak{F}(q,\mathbf{Y})$ (which is equivalent to minimizing $KL(q||p)$) over that class. The functional $\mathfrak{F}(q,\mathbf{Y})$ is referred to as the evidence lower bound (ELBO). In the case of mean-field VB, the restriction of $q(\vtheta)$ is to a product form $q(\vtheta) = \prod_{j=1}^{J}q_{j}(\vtheta_{j})$. In the context of our model we assume
\begin{equation}
\label{VB_approx}
	P(\vTheta | Y) \approx \Big[ \prod_{i = 1}^{d}  q(\vW_{(i)})q(\omega^2_i) \Big] q(\boldsymbol{\Sigma})
\end{equation}
where $\vW_{(i)}$ is the $i_{th}$ row $\vW$.

\begin{algorithm}[H]
\caption{Gibbs Sampling Algorithm}
\begin{enumerate}
     \small
\item Set tuning parameters $\lambda^{2}$ and $\rho$.
\item Initialize $\boldsymbol{W},\boldsymbol{\Sigma},\boldsymbol{\omega^2}$ and repeat steps (3) - (6) below to obtain the desired Monte Carlo sample size after burn-in. 
\item Let $\boldsymbol{W}_{(i)}$ be the $i_{th}$ row of $\mathbf{W}$,  $\boldsymbol{W}_{(-i)}$ be the rest after removing   $\boldsymbol{W}_{(i)}$ from $\boldsymbol{W}$. For $i = 1,..., d,$ update $\boldsymbol{W}_{(i)}^{T}$ as:
\begin{equation*} {\boldsymbol{W}_{(i)}^{T}} \sim \text{MVN}_{c} ( \; \bvmui ,\; \boldsymbol{\Sigma}_i ),
\vspace{-1em}
\end{equation*}

\item Where:
\begin{equation*}
\begin{split}
 \bvmui &=\boldsymbol{\Sigma}_i \left(- \sumln (\boldsymbol{x}_{\ell {(i)}} \otimes I_c)[(D_A-\rho A)\otimes\boldsymbol{\Sigma}^{-1}]\right.\times (\boldsymbol{x}_{\ell (-i)}^{T} \otimes I_c) (\boldsymbol{W}_{(-i)}^T) \\
 & \left. + \sumln (\boldsymbol{x}_{\ell {(i)}} \otimes I_c)[(D_A-\rho A)\otimes\boldsymbol{\Sigma}^{-1}] \boldsymbol{y}_{\ell} \right),
\end{split}
\end{equation*}
\begin{equation*}
\begin{split}
\boldsymbol{\Sigma}_i &= \left(\boldsymbol{H}_i+\sumln (\boldsymbol{x}_{\ell {(i)}} \otimes I_c)[(D_A-\rho A)\otimes\boldsymbol{\Sigma}^{-1}] \right. \left. (\boldsymbol{x}_{\ell (i)}^{T} \otimes I_c) \right)^{-1},
\end{split}
\end{equation*}
\begin{equation*}
\boldsymbol{H}_i = \left[ \frac{1}{{\omega_i^2}} \otimes I_{\frac{c}{2}}\otimes \Sigma^{-1}\right].
\end{equation*}
 
\item Update $\boldsymbol{\Sigma}$ as:
\begin{equation*}
\boldsymbol{\Sigma} \sim \text{Inverse-Wishart}(S^*,v^*)
\end{equation*}
where:
\begin{align*}
	 S^* & =\sum\limits_{l=1}^{n} \sum\limits_{i=1}^{\frac{c}{2}} \sum\limits_{j=1}^{\frac{c}{2}}b_{i,j}\tilde{y^*_{l,i}}{\tilde{y^*_{l,i}}}^T +\sum\limits_{i=1}^{d}\sum\limits_{j^{*}=1}^{\frac{c}{2}}\frac{  \tilde{W}_{i,j^{*}} {\tilde{W}_{i,j^{*}}}^T}{\omega_i^2}+S,\\
v^* &=2n+\frac{cd}{2}+v,
\;\;\boldsymbol{y}^*_l =\boldsymbol{y}_l-\boldsymbol{W}_{}^T\boldsymbol{x}_l,\\
 {\tilde{y^*}_{l,j^*}}^{T}&=(y^*_{l,j},\;y^*_{l,j+1}), 
 \;\;\tilde {W}_{i,j^{*}}=({W_{i,j}},\;{W_{i,{j+1}}}).
\end{align*}

\item For $i = 1, \dots,d$ update $\omega_{i}^{2}$, through
\begin{equation*}
1/\omega_{i}^{2} \sim \text{Inverse-Gaussian} \left( \sqrt{ \frac{\lambda^2}{c_i^*}} \;, \;\;\lambda^2 \right)
\end{equation*}
where:
\begin{equation*}
c_i^{*}=tr(  \sum\limits_{j^{*}=1}^{\frac{c}{2}}{  \tilde{W}_{i,j^{*}} {\tilde{W}_{i,j^{*}}}^T{\boldsymbol{\Sigma}}^{-1}})
\end{equation*}
\end{enumerate}
\end{algorithm}

\begin{algorithm}[H]
   \small
\caption{Mean-field Variational Bayes Algorithm}
\begin{enumerate}
\item Set tuning parameters $\lambda^{2}$ and $\rho$ and convergence parameters $K$ and $\epsilon$.
\item Initialize $q_{(\boldsymbol{W})},q_{(\boldsymbol{\Sigma})},q_{(\boldsymbol{\omega^2})}$ and cycle through steps (3) - (5) below until the absolute relative change in the evidence lower bound $\mathcal{L}(q)$ (ELBO) is smaller than $\epsilon$ for $K$ iterations.
\item For $i = 1,..., d,$ update \\
\begin{equation*}
\begin{split}
&\vSigma_{q(\boldsymbol{W}_{(i)})}^{-1} \gets \bigg(  \left[   \mu_q({\eta_i})  \otimes I_{\frac{c}{2}}\otimes (v_{q(\boldsymbol{\Sigma}) }S_{q(\boldsymbol{\Sigma})}^{-1})\right] +\sumln (\boldsymbol{x}_{\ell {(i)}} \otimes I_c)[(D_A- \rho A))\otimes(v_{q(\boldsymbol{\Sigma}) }S_{q(\boldsymbol{\Sigma})}^{-1}) ](\boldsymbol{x}_{\ell (i)}^{T} \otimes I_c) \bigg )^{-1},
\end{split}
\end{equation*}
\begin{equation*}
\begin{split}
&\vmu_{q(\boldsymbol{W}_{(i)})} \gets \vSigma_{q(\boldsymbol{W}_{(i)})} \bigg(- \sumln (\boldsymbol{x}_{\ell {(i)}} \otimes I_c)[(D_A- \rho A)\otimes (v_{q(\boldsymbol{\Sigma}) }S_{q(\boldsymbol{\Sigma})}^{-1})](\boldsymbol{x}_{\ell (-i)}^{T}
\otimes I_c) \big(\vmu_{q(\boldsymbol{W}_{(-i)})}\big) \\
&+ \sumln (\boldsymbol{x}_{\ell {(i)}} \otimes I_c)
[(D_A- \rho A)\otimes (v_{q(\boldsymbol{\Sigma}) }S_{q(\boldsymbol{\Sigma})}^{-1})] \boldsymbol{y}_{\ell} \bigg) 
\end{split}
\end{equation*}

\item  Update $S_{q(\boldsymbol{\Sigma})}$ as
\begin{equation*}
\begin{split}
&S_{q(\boldsymbol{\Sigma})} \gets \sum\limits_{l=1}^{n} \sum\limits_{i=1}^{\frac{c}{2}} \sum\limits_{j=1}^{\frac{c}{2}} b_{i,j}\tilde{y^*_{l,i}}{\tilde{y^*_{l,i}}}^T +
\sum\limits_{i=1}^{d}\sum\limits_{j^{*}=1}^{\frac{c}{2}}{ E_q\big( \tilde{W}_{i,j^{*}} {\tilde{W}_{i,j^{*}}}^T\Big)}\mu_{q(\eta_i)} +S
\end{split}
\end{equation*}
where:
	 $$b_{i,j} = \left[D_A- \rho A\right]_{i,j}$$
\item for $i = 1,...,d$, update $\mu_{q(\eta_i)}$
\begin{equation*}
\mu_{q(\eta_i)} \gets \sqrt{\frac{\lambda^2}{E_q(c_i^*)}}  
\end{equation*}
	  where:

      $$E_q(c_i^*) = E_q \left(tr(  \sum\limits_{j^{*}=1}^{\frac{c}{2}}{  \tilde{W}_{i,j^{*}} {\tilde{W}_{i,j^{*}}}^T\Sigma^{-1}})\right)$$
Update:
	    $$\mu_{q(\omega_i^2) } \gets  \frac{1}{\mu_{q(\eta_i)}} + \frac{1}{\lambda^2 }$$
	    $$Var_{q(\omega_i^2)}  \gets \frac{1}{\mu_{q(\eta_i)}\lambda^2} + \frac{2}{(\lambda^2)^2}$$

\end{enumerate}
\end{algorithm}
 
 We maximize the functional $\mathfrak{F}(q_{1},\dots,q_{J},\mathbf{Y})$ over the $q_{j}$'s using a coordinate ascent procedure. The update steps for this procedure take the form (see, e.g., Ormerod and Wand, 2010)
    $$
    q_{i}(\vtheta_{i}) = \frac{\exp\{ E_{\theta_{-i}}[\log P(\vtheta_{i}|\mathbf{Y},\vtheta_{-i})]\}}{\int\exp\{ E_{\theta_{-i}}[\log P(\vtheta_{i}|\mathbf{Y},\vtheta_{-i})]\} d\vtheta_{i}}
    $$
where the expectation is taken with respect to  $q_{-i}(\vtheta_{-i}) = \prod_{l \ne i} q_{l}(\vtheta_{l})$. This leads to a set of update equations that are iterated until  convergence to a local optimum. These update equations are presented in Algorithm 2 and their derivations are detailed in the Supplementary Material (Web Appendix B). On convergence, the approximation to the posterior distribution is based on (\ref{VB_approx}) as well as the solutions
 \begin{align*}
	q(\vW_{(i)}) &\equiv \text{MVN}(\vmu_{q_{(\boldsymbol{W}_{(i)})}},  \vSigma_{q_{(\boldsymbol{W}_{(i)})}}), \,\, i=1, \dots, d,\\
	q(\omega^2_i) &\equiv \text{Reciprocal Inverse Gaussian}(\mu_{q(\eta_i)} , \lambda_{q(\eta_i)} ), \,\, i=1, \dots, d,\\
	q(\vSigma) &\equiv  \text{Inverse-Wishart}(S_{q(\boldsymbol{\Sigma})}, v_{q(\boldsymbol{\Sigma})})
	\end{align*}
where the statistics  $\{\vmu_{q_{(\boldsymbol{W}_{(i)})}},  \vSigma_{q_{(\boldsymbol{W}_{(i)})}}, \, i=1,\dots, d$\}; $\{\mu_{q(\eta_i)}, \lambda_{q(\eta_i)}, i=1,\dots, d\}$; $S_{q(\boldsymbol{\Sigma})}$, $v_{q(\boldsymbol{\Sigma})}$, also referred to as variational parameters, are obtained as the output of the iterative Algorithm 2.

To initialize the variational Bayes algorithm, we use a ridge regression estimator obtained separately for each column of $\boldsymbol{W}$ obtained by fitting ridge regression with individual scalar-valued phenotypes as the response. The ridge estimators are then used to initialize the mean of the variational posterior distribution. The output of variational Bayes is then used to initialize the MCMC sampling algorithm.

\subsection{Bayesian FDR}
The Bayesian FDR procedure applied in our work for SNP selection follows the approach developed in Morris et al. (2008), but it has been adapted and implemented for the current spatial model. We assume that we have $N$ samples $W_{i,j}^{(1)},\dots, W_{i,j}^{(N)}$ from the posterior distribution for each of the regression coefficients $W_{i,j}$,  $i=1,\cdots, d$, $j=1,\cdots, c$. Let  $c^*$ be a known critical value that is chosen a priori to represent an effect size of interest. Given this value, we compute a posterior tail probability for the $i$-th SNP at region $j$ as 
$ p_{i,j}= Pr(|W_{i,j}|>c^*|\textbf{Y}),i=1,\dots,d; j=1,\dots,c,$
which can be approximated by 
$ p_{i,j} \approx N^{-1} \sum_{i^*=1}^{N} I\left\lbrace|W_{i,j}^{(i^*)}|>c^* \right\rbrace$
and we replace any $p_{i,j}=1$ with $1-(2N)^{-1}$. Given these posterior tail probabilities and a desired expected Bayesian FDR-bound $\alpha$, we denote by $\phi_\alpha$ the corresponding threshold chosen so that a SNP-region pair $(i,j)$ is selected if $p_{i,j} > \phi_{\alpha}$. The cut-off $\phi_\alpha$ can be computed by sorting 
$\left\lbrace p_{i,j} ,i=1,\cdots,d; j=1,\dots,c \right\rbrace$ in descending order $\left\lbrace p(i) ,i=1,\cdots,d \times c \right\rbrace$, then $\phi_\alpha=p(\lambda)$, with 
\begin{equation}
\label{posterior_FDR}
\lambda=\text{max}\left\lbrace l^* : (l^*)^{-1}\sum_{l=1}^{l^*}\left(1-p(l)\right)\leq\alpha \right\rbrace. 
\end{equation}
The threshold $\phi_\alpha$ is a cutpoint on the posterior probabilities that controls the expected Bayesian FDR below level $\alpha$. The value of $c^*$ can be chosen based on prior knowledge of what constitutes an effect size of interest, or in the absence of such knowledge, it can be chosen based on the data.

\subsection{Model Selection and Tuning}

To compare the spatial and non-spatial models and to choose values for the tuning parameters, one option is the use of the WAIC (Vehtari et al., 2017). This criterion can be computed from posterior simulation output and takes the form  
$$
WAIC = -2\sum_{l=1}^{n}\log E_{\W,\vSigma}[p(\yl | \W,\vSigma) | \mathbf{y}_{1}, \dots, \mathbf{y}_{n}] 
$$
$$
+ 2\sum_{l=1}^{n} VAR_{\W,\vSigma}[\log p(\yl | \W,\vSigma) | \mathbf{y}_{1}, \dots, \mathbf{y}_{n}]
$$
where $p(\yl | \W,\vSigma)$ is the multivariate normal density function associated with the conditional autoregressive model (\ref{CAR_MVN}), and the expectation and variance are taken with respect to the posterior distribution, with lower values being preferred. 

An alternative approach to tuning the model is to use a simple modified moment estimator based on the ridge regression estimator that we use to initialize VB. This estimator implicitly uses cross-validation (CV) which is used to tune the ridge estimator and is much faster than applying CV directly to our Bayesian model. The estimator takes the form
$$
\hat{\lambda}^{2} = \frac{dc(c+1)}{\max\{1, \nu-3\}}  \left(\sum_{i,j}\hat{\mathbf{W}}_{R_{i,j}}^{2}\right)^{-1},
$$
and its form is derived in Web Appendix C. 

For VB we use the moment estimate for $\lambda^{2}$ and fix the value of $\rho = 0.95$ corresponding to a reasonable degree of spatial dependence on the graph represented by the neighbourhood matrix $\boldsymbol{A}$. The MCMC algorithm is then initialized using the output of VB and the value of $\lambda^{2}$ can be set at the moment estimate while $\rho$ can be chosen using the WAIC. The WAIC can also be used to compare the spatial model with the non-spatial model. 

While the WAIC can be used for selection of $\rho$, we have found that selection of $\lambda^{2}$ using the WAIC can result in patterns where WAIC decreases in a monotone fashion as $\lambda^{2}$ increases. In these situations, we recommend that either the modified moment estimate (which is itself based on CV) is used or that the results be summarized over a range of values for $\lambda^{2}$ through the regularization path. 

\section{Simulation Studies}

We conduct a simulation study where the dimension of the data is the same as that in our motivating application with 56 structural brain imaging phenotypes, 486 SNPs from 33 genes, and 632 subjects. The data are simulated from the spatial model with parameter values set at the estimates obtained from the real data. 

We use 100 simulation replicates and run the MCMC algorithms for both the spatial model and the non-spatial model for 10000 iterations with 5000 iterations used as burn-in for each replicate, and we run the variational Bayes algorithm to convergence of the ELBO, where convergence is declared when the absolute relative change in the ELBO is smaller than $\epsilon = 10^{-4}$ for $K=2$ consecutive iterations. The tuning parameters are chosen using the WAIC over a coarse grid for the spatial model implemented via MCMC; set using the default approach in the R package for the model of Greenlaw et al. (based on five-fold cross-validation over a grid of possible values with out-of-sample prediction based on an approximate posterior mode, the estimator of Wang et al. (2012), that can be computed quickly) and using our moment based approach to obtain $\lambda^{2}$ for VB, while the spatial parameter is held fixed at the default value of $\rho = 0.95$ for VB (the true value set when simulating the data is $\rho = 0.8$).

Five measures are used for comparison in this study and are listed in Table 1. For the best overall estimation performance, the spatial model implemented using MCMC has an average mean-squared error (MSE) of 0.0055, followed by the non-spatial model with an average MSE of 0.012 and finally the spatial model implemented with VB with an average MSE of 0.043. Here, the average is taken over the 27,216 regression coefficients in the model. Next, we examine the correlation between the posterior mean estimates of $\text{vec}(\mathbf{W})$ and the true values averaged over simulation replicates. The spatial model implemented using MCMC has the best performance (0.69), followed by the non-spatial model implemented via MCMC (0.67) and then the spatial model implemented using VB (0.65). The average-squared bias is lowest for the non-spatial model (3.29$\times 10^{-4}$), followed by the spatial model implemented with MCMC (4.70$\times 10^{-3}$) and finally the spatial model implemented using VB (9.10$\times 10^{-3}$) has the highest average-squared bias. Both the non-spatial and spatial models appear to exhibit adequate coverage probability overall when implemented with MCMC while the variational Bayes implementation of the spatial model exhibits coverage of 95\% equal-tail credible intervals that is slightly lower at 0.91. The average posterior standard deviation obtained from the spatial model (MCMC) is slightly larger than that obtained from the non-spatial model perhaps indicating that the spatial model is adequately accounting for the dependence. As is typical with variational Bayes the posterior standard deviation is underestimated. This arises from the Kullback-Leibler objective function $KL(q||p)$ for variational Bayes which under-penalizes approximations that are under-dispersed.

To evaluate the empirical FDR of the Bayesian FDR approach when the expected Bayesian FDR is controlled with $\alpha = 0.05$, leading to the threshold for posterior probabilities $\phi_{\alpha=0.05}$, we conduct an additional 100 simulation replicates where the error structure underlying the simulated data is again set to that obtained from the real data application. All but 100 SNPs have their 56 coefficients set to exactly zero while 50 rows of {\bf W} have coefficients set to $W_{ij}=1$, 25 rows have coefficients set to $W_{ij} = 2$ and the remaining 25 rows have coefficients set to $W_{ij} = 3$. Here we evaluate the empirical FDR for the spatial model and both of its implementations. The methodology of Greenlaw et al. (2017) uses credible intervals rather than Bayesian FDR for SNP selection. Both the VB and MCMC algorithms use the moment estimate for $\lambda^{2}$. The empirical proportion of false discoveries averaged over simulation replicates is reported in Supplementary Material Table 2 for both VB and MCMC as a function of $c^{*}$. We note that these $c^{*}$ values are based on each simulated response matrix {\bf Y} having its columns centered and scaled. The empirical FDR appears to drop to 0 with increasing $c^{*}$ faster with the MCMC implementation which is consistent with the VB implementation being more liberal in selecting SNPs. 

\section{ADNI-1 Study of MRI and Genetics}

We apply our spatial model as well as the group sparse multi-task regression model of Greenlaw et al. (2017) to MRI and genetic data collected from $n = 632$ subjects from the ADNI-1 database. The genetic data comprise SNPs belonging to the top 40 Alzheimer's Disease (AD) candidate genes listed on the AlzGene database as of June 10, 2010. The data presented here are queried from the genome build as of December 2014, from the ADNI-1 data. After quality control and imputation steps, the genetic data used for this study include 486 SNPs from the 33 targeted genes discussed in  Szefer et al. (2017). The freely available software package PLINK (Purcell et.al., 2007) is used for genomic quality control.  Subjects are included if their genotyping data is available, they have a baseline MRI scan, and they have at least one additional follow-up baseline scan. Among all SNPs, only SNPs belonging to the top 40 AD candidate genes listed on the AlzGene database (www.alzgene.org) as of June 10, 2010, are selected after the standard quality control (QC) and imputation steps. The QC criteria for the SNP data include (i) call rate check per subject and per SNP marker, (ii) gender check, (iii) sibling pair identification, (iv) the Hardy-Weinberg equilibrium test, (v) marker removal by the minor allele frequency and (vi) population stratification. Our thresholds for SNP and subject exclusion are the same as in Wang et al. (2012) with three exceptions. In the data quality control step, we used a stricter minimum call rate of 95\% on SNPs vs. Wang et al.'s call rate of 90\%. To assign SNPs to genes, we use a genome build (Build GRCh38.p2) from December 2014 whereas these authors use a genome build (Build 36.2) from September 2006, and use all subjects with a baseline measurement whereas we choose subjects with a baseline MRI scan and a scan at at least one additional time point in the longitudinal study. 

The response measures are obtained by preprocessing the MRI data using the FreeSurfer V4 software which conducts automated parcellation to define volumetric and cortical thickness values from the $28$ ROIs considered in Szefer et al. (2017) and Greenlaw et al. (2017) on each hemisphere of the brain, leading to $c=56$ brain measures in total. These ROIs are chosen based on prior knowledge that they are related to Alzheimer's Disease. Each of the response variables are adjusted for age, gender, education, handedness, baseline total intracranial volume (ICV), potential population stratification and APOE genotype and centered to have zero-sample-mean and unit-sample-variance. 

We fit our new spatial model to these data using both Algorithm 1 (Gibbs sampling) and Algorithm 2 (VB). In addition, we fit the non-spatial model of Greenlaw et al. (2017) using the MCMC sampler derived therein, with the tuning parameters for the non-spatial model set at $\lambda_{1}^{2}=1000$ and $\lambda_{2}^{2}=1000$ based on the values selected in Greenlaw et al. (2017). In all cases, MCMC sampling is run for 10,000 iterations with the initial 5,000 iterations discarded. The required computation time for the spatial model (MCMC) is 50 hours on a single core (2.66-GHz Xeon x5650) with 20GB of RAM, while the computation for the non-spatial model is 5hrs. Some trace plots and MCMC convergence diagnostics are presented in Supplementary Material Web Appendix E and these demonstrate rapid convergence and good mixing of the MCMC sampling chains. The VB algorithm is run to convergence and requires 45 minutes with the ELBO converging in approximately 16 iterations. The tuning parameter $\lambda^{2}$ is set based on the moment estimator and we set $\rho = 0.95$ for VB.  The convergence of VB based on successive values of the ELBO is depicted in Figure 2.

\begin{figure}[htp]
\centering
\caption{Variational Bayes - convergence of the evidence lower bound (ELBO) for the ADNI MRI and genetic data considered in the application.}	
\includegraphics[scale=0.3]{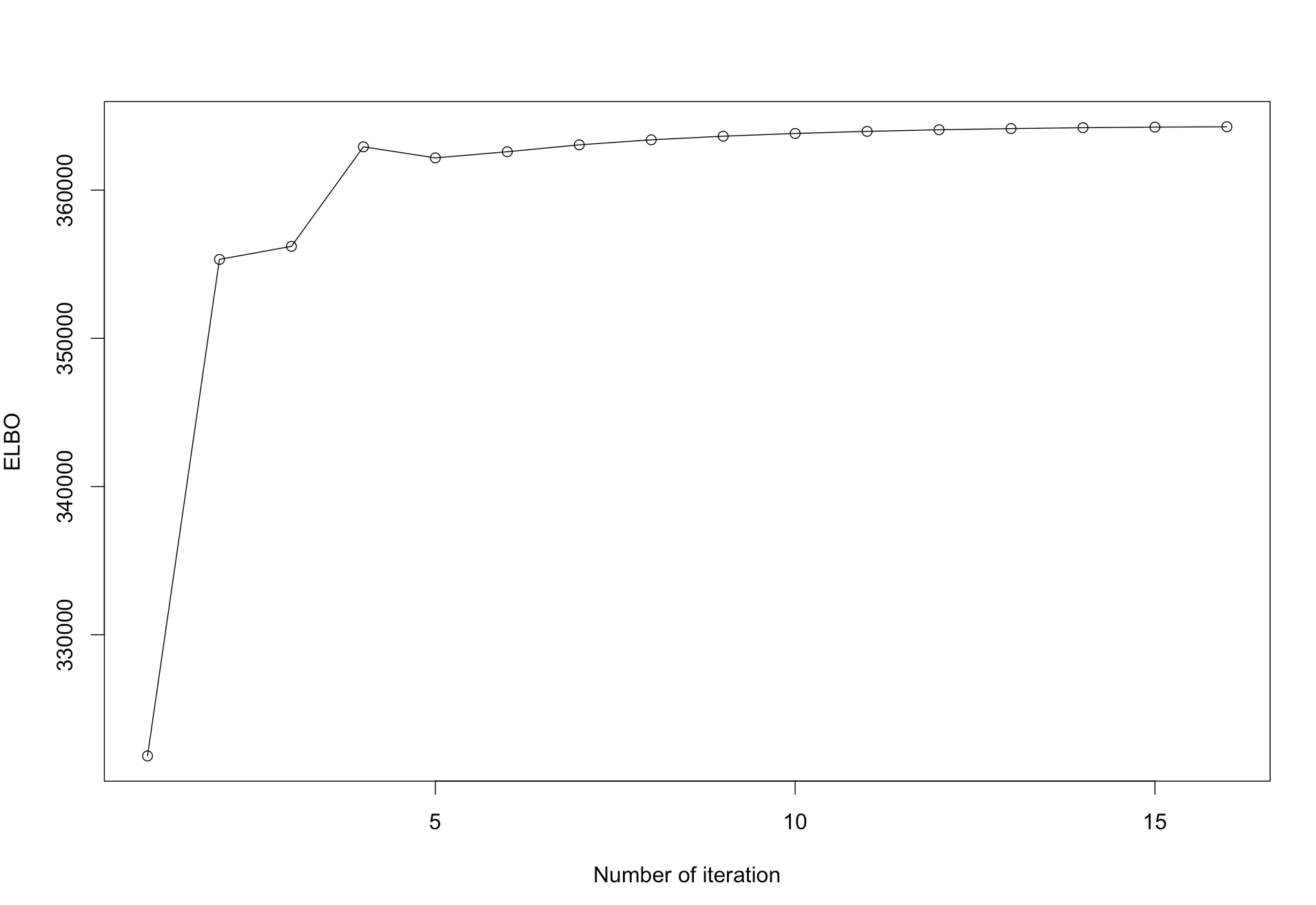}
\label{snpregion}
\end{figure}

Supplementary Material Figure 3 presents the WAIC computed for a number of different choices of the tuning parameters $\rho$ and $\lambda^{2}$. For the values of $\rho=0.8$ and $\lambda^{2} = 10,000$, the value of the WAIC is $83,170$. While the WAIC is able to identify a value for $\rho$, it is monotone decreasing as a function of $\lambda^{2}$. We will thus summarize the results over several values of $\lambda^{2}$. The WAIC obtained for the non-spatial model with value of the tuning parameters used in Greenlaw et al. (2017) ($\lambda_{1}^{2}=1000$ and $\lambda_{2}^{2}=1000$) is $108,745$. 


Figure 3 presents the number of SNPs chosen by the spatial model for each ROI using Bayesian FDR (based on a critical value of $c^{*} = 0.044$) as a function of the tuning parameter $\lambda^{2}$ for both Gibbs sampling and VB. As expected, the curves are monotone decreasing but we note that their shapes differ when comparing the algorithms. In particular, VB selects a larger number of SNPs at all values of $\lambda^{2}$. We suggest that the VB algorithm be used for obtaining starting values to initialize the MCMC as well as a tool to gain some initial insight into the data while the MCMC sampler runs to completion. This is useful because the MCMC sampler requires a relatively long run time, and the VB algorithm can be used initially (requiring 45 minutes in our study) while the MCMC sampler runs (requiring 50 hrs in our study).

\begin{figure}[h]
\centering
\caption{ADNI-1 Data - Relationship between the number of selected SNPs in each region and $\lambda^2$. Each region is represented with a curve in each panel of the figure. The left panel shows this relationship for MCMC combined with Bayesian FDR ($\alpha = 0.05$) while the right panel shows the same relationship for VB with Bayesian FDR ($\alpha = 0.05$).}
\includegraphics[scale=0.15]{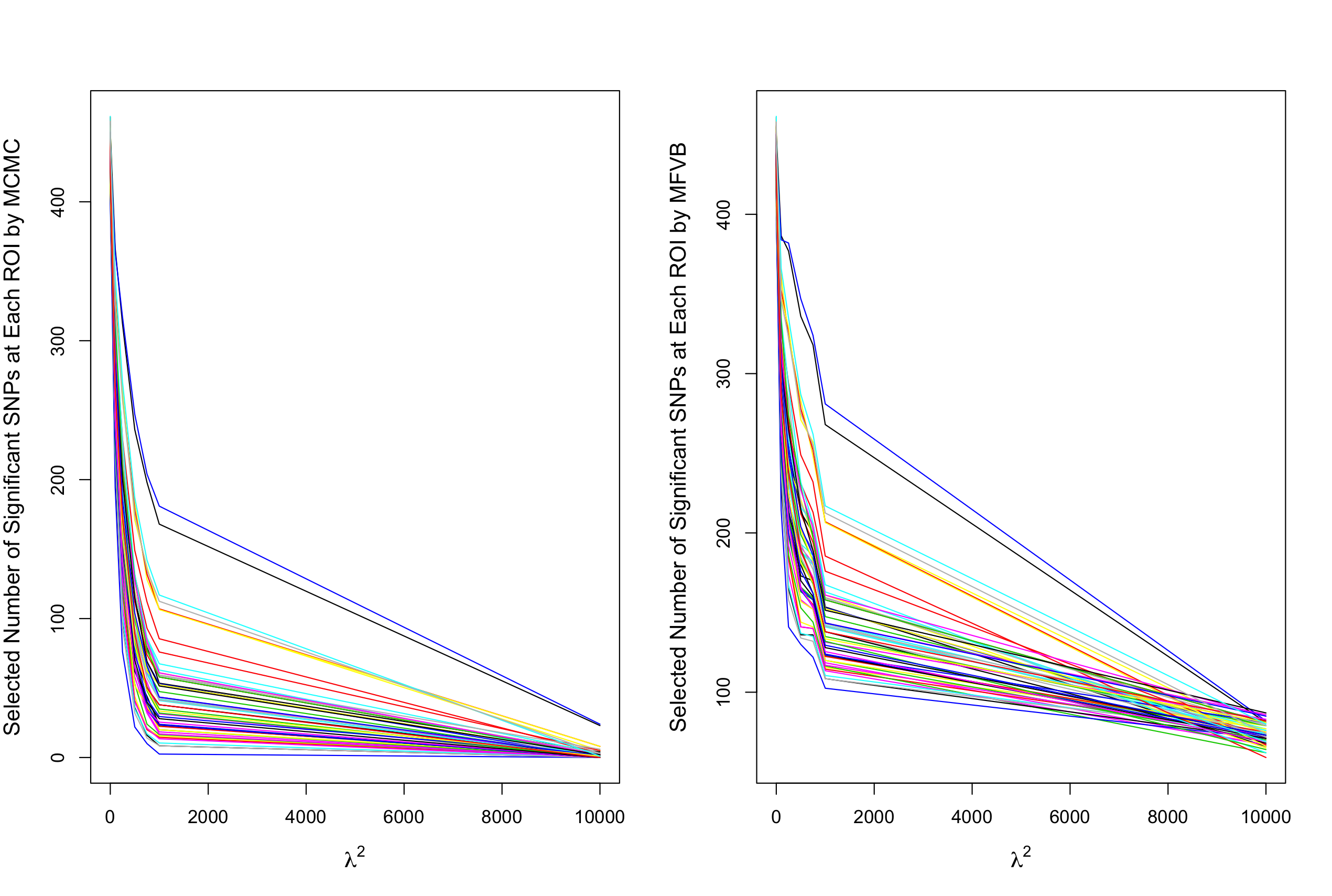}
\label{waicf}
\end{figure}

For the values of the tuning parameters $\rho=0.8$, $\lambda^{2}=10,000$, the average number of SNPs selected per ROI is 2, while more than half of the ROIs have no SNPs selected. In total, 75 SNPs across all 56 ROIs are selected and these are listed in Table 1 of the Supplementary Material along with the corresponding phenotypes that they are associated with. With the VB approximation, 150 SNPs are selected, and the set of SNP-ROI pairs selected by MCMC is a proper subset of the set selected by VB. In addition, the subset of SNPs and phenotypes also selected by the approach of Greenlaw et al. (2017) where the marginal posterior 95\% credible interval is used for SNP selection are also highlighted in bold in Table 1 of the Supplementary Material.

Considering all three approaches, a consistent signal is found at the APOE gene, where all three methods select SNP rs405509 and the Bayesian FDR procedure selects this SNP for four phenotypes, namely, the middle temporal gyrus thickness on the right hemisphere, supramarginal gyrus thickness on the right hemisphere, mean thickness of the caudal midfrontal, rostral midfrontal, superior frontal, lateral orbitofrontal, and medial orbitofrontal gyri and frontal pole on the right hemisphere, and the mean thickness of the inferior temporal, middle temporal, and superior temporal gyri on the right hemisphere. We note that the selected associations for this SNP all correspond to ROIs in the right brain hemisphere. The associations between the genetic signal represented in our analysis by APOE SNP rs405509 with phenotypes on the right hemisphere of the brain may be of potential interest for further investigation.

The associated point estimates and 95\% equal-tail credible intervals for all 56 phenotypes and APOE SNP rs405509 are presented in Table 3 of the Supplementary Material and a subset of these results for 26 phenotypes is presented in Table 2. Overall, there is broad agreement based on the overlap of the interval estimates, though, we also see a number of examples where the 95\% interval estimate obtained from the spatial model includes the value 0 while the corresponding interval estimate for the non-spatial model does not include zero. The higher bias arising from VB in the simulation studies may also be apparent in the results presented in Table 2. The interval estimates arising from VB appear to be slightly more narrow than those obtained from MCMC, but not to a large degree, and this may also be reflected in the selection rates depicted in Figure 3 which are slightly higher. This is consistent with the coverage probabilities found in the first simulation study.

Another consistent signal is found at the ACE gene with SNP rs4311, which is found associated with 12 ROIs. We note that all but one of these ROIs is in the right hemisphere, and three of these ROIs (all of which are in the right hemisphere) are in common with the ROIs selected for this SNP by Greenlaw et al. (2017). In Figure 4 we indicate the SNPs chosen for each ROI, where the SNPs are grouped on the x-axis by gene and the ROIs are grouped in left/right pairs on the y-axis. The selected SNPs for each ROI are shown for tuning parameter values $\lambda^{2}=1000$ and $\lambda^{2} = 10,000$. In both cases the value of the spatial tuning parameter for the CAR model is set at $\rho = 0.8$ as suggested by Supplementary Material Figure 3.

\begin{figure}[htp]
\centering
\caption{ADNI-1 Data: SNPs chosen with the spatial model fit using Gibbs sampling and Bayesian FDR ($\alpha = 0.05$) are highlighted in red for each phenotype. The black ticks on y-axis indicate the phenotypes from the left/right hemisphere, and the SNPs from same gene are indicated by the ticks on x-axis. The top panel corresponds to the case $\lambda^{2}=1000$ while the bottom panel corresponds to the case $\lambda^{2}=10,000$.}
\label{sigsnpperregion}
\includegraphics[scale=0.35]{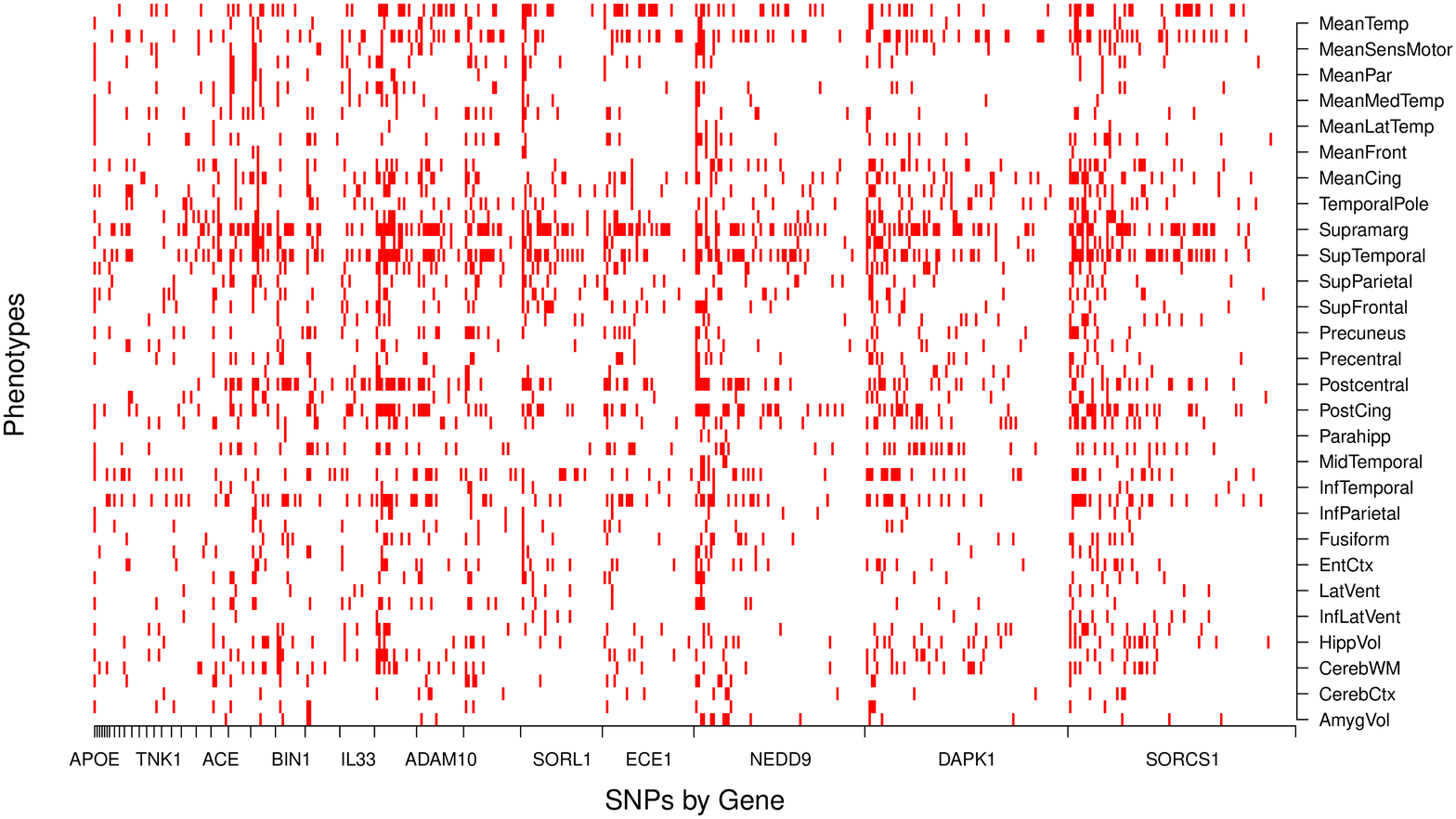}
\includegraphics[scale=0.35]{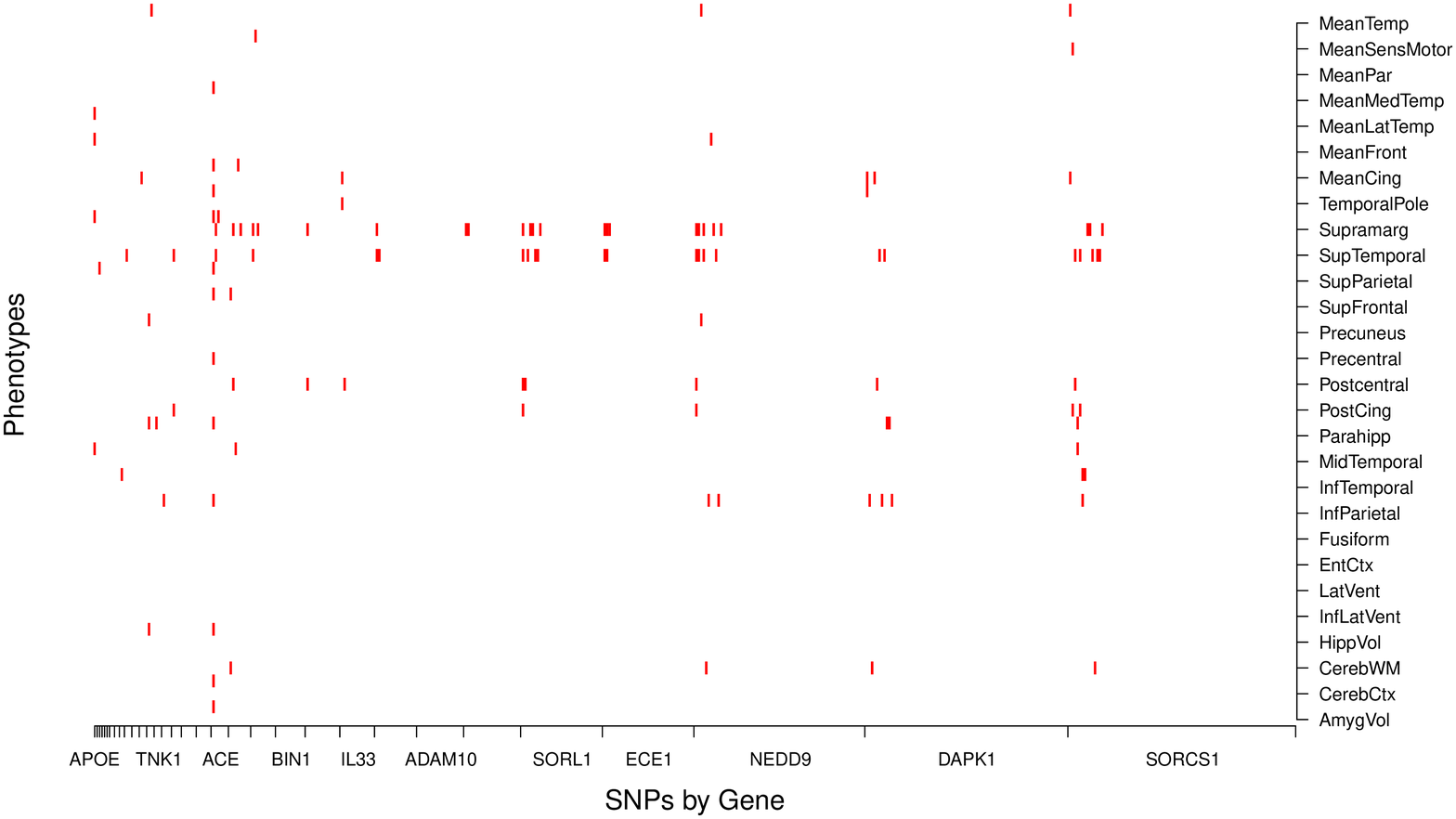}
\end{figure}

Examining Figure 4, two ROIs stand out has having a relatively broad genetic signal that persists even as the tuning parameter increases from $\lambda^{2} = 1000$ to $\lambda^{2} = 10,000$. These are \emph{Left-Supramarg} (thickness of the left supramarginal gyrus) and \emph{Left-SupTemporal} (thickness of the left superior temporal gyrus). For the case where $\lambda^{2} = 1000$, phenotype \emph{Left-Supramarg} is associated with 188 SNPs (top panel of Figure 4) and this decreases to 24 SNPs (bottom panel of Figure 4) when $\lambda^{2} = 10,000$. When $\lambda^{2} = 1000$ phenotype \emph{Left-SupTemporal} is associated with 188 SNPs and this decreases to 23 SNPs when $\lambda^{2} = 10,000$. This is to be compared with the average number of SNPs selected over all ROIs when $\lambda^{2} = 10,000$ which is just 2. The regularization paths for these two regions are shown in Supplementary Material Figure 4 where the SNPs having the most persistent signal across values of $\lambda^{2}$ are highlighted. These include SNP rs10868609 for the thickness of the left supramarginal gyrus and rs10501426 for the thickness of the left superior temporal gyrus.


\section{Conclusion}

We have developed a spatial multi-task regression model for relating genetic data to imaging phenotypes. The error structure for the imaging phenotype is based on a computationally convenient proper bivariate conditional autoregressive model, which allows us to account for bilateral correlation across brain hemispheres while also allowing us to account for within-hemisphere correlation using a graph structure. Examination of the data can assist in the determination of whether a model allowing for correlation is required. In our case it is clear that a model accounting for correlation is indeed required (see, e.g., Figure 1) for bilateral correlation. Empirical correlation between neighbouring regions can also be computed to determine if a simpler model should be considered. Examination of alternative spatial models and comparisons with respect to estimation, inference and robustness relative to that currently implemented under different settings will be part of future work.  

Future developments may also incorporate the Bayesian false discovery probability proposed by Wakefield et al. (2007) which allows the user to account explicitly for the cost of false discoveries and the cost of false non-discovery. With regards to the two computational algorithms, we recommend that the approximate VB procedures be used to initialize the MCMC algorithm and also to obtain an initial insight into the data while the MCMC sampler runs. It appears that VB combined with Bayesian FDR tends to be more liberal in the selection of SNPs, and in our application the SNP-ROI pairs selected by MCMC + Bayesian FDR are a proper subset of that selected by VB + Bayesian FDR. Nevertheless, Figure 3 suggests that the approximation can be reasonable when good initializations are used. 


\section*{Acknowledgements}
Research is supported by funding from the Natural Sciences and Engineering Research Council of Canada and the Canadian Statistical Sciences Institute. F.S. Nathoo holds a Tier II Canada Research Chair in Biostatistics for Spatial and High-Dimensional Data. Research was enabled in part by support provided by WestGrid (www.westgrid.ca) and Compute Canada (www.computecanada.ca). 
Data collection and sharing for this project was funded by the Alzheimer's Disease Neuroimaging Initiative (ADNI) (National Institutes of Health Grant U01 AG024904) and DOD ADNI (Department of Defense award number W81XWH-12-2-0012).

\pagebreak
\newpage

\captionsetup{width=16cm}
\begin{center}
\begin{longtable}{p{40mm}l l l l l c}
\caption{ {Summaries from Simulation Study I. The table entries are average mean-squared error (MSE), average correlation (Corr.) between the posterior mean and the true value, average squared-bias (Bias$^{2}$), average coverage probability (Cov. Prob.) of 95\% equal-tail credible intervals, average posterior standard deviation (P. SD). The averages are taken over the 27,216 regression coefficients in all but 'Corr.' where the average of $Corr(vec(\hat{\mathbf{W}}),vec(\mathbf{W}_{true}))$ is taken over 100 simulation replicates and $\hat{\mathbf{W}} = E[\mathbf{W}|\bf{Y}]$.} }
\label{Sim1_Results} \\
\hline  
Model & MSE & Corr. &Bias$^{2}$ &Cov. Prob. &P. SD.\\
\hline
\endfirsthead
\multicolumn{7}{l}%
{\tablename\ \thetable\ -- \textit{Continued from previous page}} \\
\hline
 
Model & MSE & Corr. &Bias$^{2}$ &Cov. Prob&P. SD.\\
\hline
\hline
\endhead
\hline \multicolumn{5}{r}{\textit{Continued on next page}} \\
\endfoot
\hline
\endlastfoot
Non-Spatial MCMC &0.012 &0.67 &3.29$\times 10^{-4}$& 0.96& 0.087\\
Spatial Model MCMC &0.0055 &0.69 &4.70$\times 10^{-3}$& 0.97& 0.10\\
Spatial Model VB & 0.043& 0.65 & 9.10$\times 10^{-3}$& 0.91& 0.037\\
\end{longtable}
\end{center}
\normalsize

\captionsetup{width=15cm}
\small
\begin{center}
\begin{longtable}{p{30mm}llllll}
\caption{ {ADNI-1 Study: posterior means and $95\%$ equal-tail credible intervals for a subset of the ROIs and their association with APOE SNP rs405509.} }
\label{APOECI} \\
\hline
  & \multicolumn{2}{c}{ \textbf{Spatial Model (MCMC)}}
  & \multicolumn{2}{c}{ \textbf{Spatial Model (MFVB)}}
 &\multicolumn{2}{c}{ \textbf{Non-Spatial Model (MCMC)}} \\
  
 Region & Mean & $95\%$ CI & Mean & $95\%$ CI & Mean & $95\%$ CI \\ 
\hline
\endfirsthead
\multicolumn{7}{l}%
{\tablename\ \thetable\ -- \textit{Continued from previous page}} \\
\hline
   & \multicolumn{2}{c}{ \textbf{Spatial Model (MCMC)}}
  & \multicolumn{2}{c}{ \textbf{Spatial Model (MFVB)}}
 &\multicolumn{2}{c}{ \textbf{Non-Spatial Model (MCMC)}}\\
 
 Region & Mean & $95\%$ CI  & Mean & $95\%$ CI & Mean & $95\%$ CI \\ 
\hline
\hline
\endhead
\hline \multicolumn{5}{r}{\textit{Continued on next page}} \\
\endfoot
\hline
\endlastfoot
 Amygdala volume (L) & 0.09 & [-0.03,0.22] &  0.17& [0.08,0.28]& 0.12 & [0.02,0.23] \\ 
    Cerebral white matter volume (L) & 0.09 & [-0.04,0.22] &  0.19 & [0.10,0.28]& 0.13 & [0.03,0.23] \\ 	
   Inferior lateral ventricle volume (L) & -0.13 & [-0.24,-0.01] &-0.14&[-0.24,-0.05]& -0.08 & [-0.18,0.02]\\
  Inferior parietal gyrus thickness (R) & 0.11 & [0.01,0.21] &0.20&[0.11,0.29]& 0.13 & [0.03,0.23] \\ 
  Fusiform, parahippocampal, and lingual gyri, temporal pole and transverse temporal pole mean thickness (L) & 0.12 & [0.01,0.24] &0.22&[0.13,0.32]& 0.14 & [0.04,0.24] \\ 
  Inferior temporal, middle temporal, superior temporal, fusiform, parahippocampal, and lingual gyri, temporal pole and transverse temporal pole mean thickness (L) & 0.11 & [0.01,0.22] &0.20&[0.11,0.28]& 0.13 & [0.03,0.23] \\ 
   Postcentral gyrus thickness (L) & 0.13 & [0.03,0.24] &0.23&[0.13,0.33]& 0.14 & [0.03,0.23] \\ 
  Superior frontal gyrus thickness (R) & 0.12 & [0.02,0.22] &0.19&[0.11,0.28]& 0.14 & [0.04,0.24] \\ 
  Supramarginal gyrus thickness (L) & 0.13 & [0.02,0.23] &0.21&[0.13,0.29]& 0.14 & [0.04,0.25] \\ 
  Supramarginal gyrus thickness (R) & 0.14 & [0.04,0.24] &0.24&[0.17,0.33]& 0.15 & [0.05,0.25] \\ 
  Fusiform gyrus thickness (L) & 0.11 & [-0.02,0.23] & 0.19& [0.09,0.28]& 0.13 & [0.03,0.23] \\ 
  Hippocampus volume (L) & 0.12 & [-0.02,0.26] &0.22 &[0.14,0.34]& 0.15 & [0.05,0.26] \\ 
  Caudal midfrontal, rostral midfrontal, superior frontal, lateral orbitofrontal, and medial orbitofrontal gyri and frontal pole mean thickness (L) & 0.10 & [-0.02,0.22] &0.18&[0.10,0.27]& 0.13 & [0.03,0.23] \\ 
  Inferior temporal, middle temporal, and superior temporal gyri mean thickness (L) & 0.10 & [-0.01,0.21] &0.17&[0.08,0.26]& 0.12 & [0.02,0.22] \\ 
  Inferior temporal, middle temporal, and superior temporal gyri mean thickness (R) & 0.10 & [-0.01,0.21] &0.20&[0.11,0.30]& 0.13 & [0.02,0.22] \\ 
  Fusiform, parahippocampal, and lingual gyri, temporal pole and transverse temporal pole mean thickness (R) & 0.10 & [-0.01,0.21] &0.20&[0.11,0.29]& 0.12 & [0.02,0.22] \\ 
  Inferior and superior parietal gyri, supramarginal gyrus, and precuneus mean thickness (L) & 0.09 & [-0.02,0.21] &0.16&[0.07,0.25]& 0.12 & [0.02,0.22] \\ 
  Precentral and postcentral gyri mean thickness (R) & 0.09 & [-0.02,0.20] &0.17&[0.09,0.26]& 0.12 & [0.01,0.22] \\
  Precentral and postcentral gyri mean thickness (L) & 0.10 & [-0.01,0.20] &0.16&[0.08,0.25]& 0.12 & [0.02,0.22] \\
  Middle temporal gyrus thickness (L) & 0.08 & [-0.03,0.19] &0.15&[0.06,0.23]& 0.11 & [0.01,0.21] \\ 
  Middle temporal gyrus thickness (R) & 0.09 & [-0.01,0.20] &0.18&[0.11,0.27]& 0.12 & [0.02,0.22] \\      
  Postcentral gyrus thickness (R) & 0.09 & [-0.01,0.20] &0.18&[0.08,0.28]& 0.11 & [0.01,0.21] \\ 
  Precentral gyrus thickness (R) & 0.08 & [-0.03,0.19] &0.15&[0.06,0.25]& 0.11 & [0.01,0.21] \\ 
  Precuneus thickness (R) & 0.09 & [-0.01,0.19] &0.17&[0.08,0.25]& 0.11 & [0.01,0.21] \\ 
  Superior parietal gyrus thickness (R) & 0.08 & [-0.02,0.18] &0.14&[0.06,0.22]& 0.11 & [0.01,0.21] \\ 
  Superior temporal gyrus thickness (R) & 0.10 & [-0.01,0.22] &0.22&[0.14,0.31]& 0.13 & [0.03,0.23] \\      
\end{longtable}
\end{center}
\normalsize

\nocite{*}
\bibliographystyle{ECA_jasa}
\bibliography{JASA_example}

\pagebreak

\begin{center}
\textbf{\large Supplementary Material for the 'A Bayesian Spatial Model for Imaging Genetics'}
\end{center}

\setcounter{equation}{0}
\setcounter{figure}{0}
\setcounter{table}{0}
\setcounter{page}{1}
\setcounter{section}{0}
\makeatletter
\renewcommand{\theequation}{S\arabic{equation}}
\renewcommand{\thefigure}{S\arabic{figure}}

\section*{Web Appendix A: Selected SNPs and the Corresponding Regions of Interest for the ADNI-1 Application}

\captionsetup{width=16cm}
\small
\begin{center}
\begin{longtable}{lll}
\caption{ {Application to ADNI-1 data: The $75$ SNPs and corresponding phenotypes selected from the proposed Bayesian spatial group lasso regression model with Gibbs Sampling combined with Bayesian FDR at $\alpha = 0.05$. These same SNP-ROI pairs are also selected by variational Bayes combined with Bayesian FDR at $\alpha = 0.05$. SNPs and phenotypes in bold correspond to those also chosen using 95\% credible intervals and the model of Greenlaw et al. (2017).}}
\label{bfdr_app_snp_new2} \\
\hline
\textbf{ SNP} & \textbf{Gene} & \textbf{Phenotype ID (hemisphere)}  \\
\hline
\endfirsthead
\multicolumn{3}{l}%
{\tablename\ \thetable\ -- \textit{Continued from previous page}} \\
\hline
\textbf{ SNP} & \textbf{Gene} & \textbf{Phenotype ID (hemisphere)}  \\
\hline
\endhead
\hline \multicolumn{3}{r}{\textit{Continued on next page}} \\
\endfoot
\hline
\endlastfoot

\textbf{rs4305} & ACE & SupTemporal(L), Supramarg(L)\\ 
\textbf{rs4311} & ACE & AmygVol(R), \textbf{CerebCtx(R)}, HippVol(R), \textbf{InfParietal(R)}, Parahipp(R), Precentral(L), \\
&&\textbf{SupFrontal(R)},\textbf{SupParietal(R)}, Supramarg(R), TemporalPole(R), MeanCing(R), \\
&&MeanMedTemp(R)\\ 
rs4353 & ACE & Supramarg(R)\\ 
\textbf{rs405509} & APOE & \textbf{MidTemporal(R)}, \textbf{Supramarg(R)}, \textbf{MeanFront(R)}, \textbf{MeanLatTemp(R)}\\ 
\textbf{rs11191692} & CALHM1 & SupTemporal(L)\\ 
\textbf{rs3811450} & CHRNB2 & SupParietal(R)\\ 
\textbf{rs2025935} & CR1 & Postcentral(L), Supramarg(L)\\ 
rs10780849 & DAPK1 & InfParietal(R)\\ 
rs1105384 & DAPK1 & TemporalPole(R), MeanCing(L)\\ 
\textbf{ rs1473180} & DAPK1 & CerebWM(L)\\ 
 \textbf{rs17399090} & DAPK1 & MeanCing(L)\\ 
 \textbf{rs3095747} & DAPK1 & Postcentral(L)\\ 
 rs3118853 & DAPK1 & SupTemporal(L)\\ 
 \textbf{rs3124237} & DAPK1 & InfParietal(R)\\ 
 rs3124238 & DAPK1 & SupTemporal(L)\\ 
 rs4877368 & DAPK1 & Parahipp(R)\\ 
 \textbf{rs4878117} & DAPK1 & Parahipp(R)\\ 
 rs913782 & DAPK1 & InfParietal(R)\\ 
 rs10916959 & ECE1 & Supramarg(L)\\ 
 \textbf{rs212539} & ECE1 & SupTemporal(L), Supramarg(L)\\ 
 rs4654916 & ECE1 & SupTemporal(L), Supramarg(L)\\ 
 \textbf{rs6584307} & ENTPD7 & InfParietal(R)\\ 
 \textbf{rs11601726} & GAB2 & SupTemporal(L), Supramarg(L)\\ 
 rs7927923 & GAB2 & SupTemporal(L)\\ 
 rs17561 & IL1A & InfTemporal(R)\\ 
 \textbf{rs16924159} & IL33 & TemporalPole(L), \textbf{MeanCing(L)}\\ 
 \textbf{rs928413} & IL33 & Postcentral(L)\\ 
 \textbf{rs1433099} & LDLR & CerebWM(L), SupFrontal(R)\\ 
 rs2228671 & LDLR & MidTemporal(R)\\ 
 \textbf{rs2569537} & LDLR & MeanCing(R)\\ 
 rs6511720 & LDLR & Postcentral(L), Supramarg(L)\\ 
 rs688 & LDLR & Supramarg(L)\\ 
 rs2184226 & MTHFR & SupTemporal(L), Supramarg(L)\\ 
 rs3737964 & MTHFR & MeanSensMotor(R)\\ 
 rs4846048 & MTHFR & Supramarg(L)\\ 
 \textbf{rs12209631} & NEDD9 & CerebWM(L)\\ 
 \textbf{rs1475345} & NEDD9 & InfParietal(R)\\ 
 rs16871157 & NEDD9 & SupTemporal(L), Supramarg(L)\\ 
 \textbf{rs17496723} & NEDD9 & MeanFront(R)\\ 
 rs2072834 & NEDD9 & Supramarg(L)\\ 
 rs2182335 & NEDD9 & Precuneus(R), MeanTemp(R)\\ 
 rs2182337 & NEDD9 & SupTemporal(L)\\ 
 rs2950 & NEDD9 & SupTemporal(L), Supramarg(L)\\ 
 rs4713379 & NEDD9 & InfParietal(R)\\ 
 \textbf{rs744970} & NEDD9 & Supramarg(L)\\ 
 rs760680 & NEDD9 & PostCing(L), Postcentral(L), SupTemporal(L), Supramarg(L)\\ 
 rs10501604 & PICALM & Supramarg(L)\\ 
 \textbf{rs7938033} & PICALM & Supramarg(L)\\ 
 rs6084833 & PRNP & PostCing(L), SupTemporal(L)\\ 
 rs10748924 & SORCS1 & InfTemporal(R)\\ 
 rs10786972 & SORCS1 & MeanCing(L), MeanTemp(R)\\ 
 \textbf{rs10787010} & SORCS1 & PostCing(L), MeanSensMotor(L)\\ 
 \textbf{rs10787011} & SORCS1 & Supramarg(L)\\ 
 rs10884399 & SORCS1 & Supramarg(L)\\ 
 rs11193198 & SORCS1 & SupTemporal(L)\\ 
 rs12240854 & SORCS1 & Postcentral(L), SupTemporal(L)\\ 
 \textbf{rs1269918} & SORCS1 & \textbf{CerebWM(L)}\\ 
 rs1887635 & SORCS1 & SupTemporal(L)\\ 
 \textbf{rs2149196} & SORCS1 & MidTemporal(R), Parahipp(R)\\ 
 rs2243581 & SORCS1 & SupTemporal(L)\\ 
 \textbf{rs2418811} & SORCS1 & PostCing(L), SupTemporal(L)\\ 
 rs596577 & SORCS1 & Supramarg(L)\\ 
 rs7903481 & SORCS1 & InfParietal(R), InfTemporal(R)\\ 
 \textbf{rs10502262} & SORL1 & Postcentral(L)\\ 
 \textbf{rs1699102} & SORL1 & PostCing(L), Postcentral(L), SupTemporal(L), Supramarg(L)\\ 
 \textbf{rs1699105} & SORL1 & SupTemporal(L)\\ 
 rs2276346 & SORL1 & Supramarg(L)\\ 
 rs3781832 & SORL1 & Supramarg(L)\\ 
 rs4936632 & SORL1 & SupTemporal(L)\\ 
 rs661057 & SORL1 & SupTemporal(L)\\ 
 rs726601 & SORL1 & Supramarg(L)\\ 
 rs762484 & TF & MeanCing(L)\\ 
 \textbf{rs1568400} & THRA & MeanTemp(R)\\ 
 \textbf{rs3744805} & THRA & HippVol(R), Parahipp(R), Precuneus(R)\\ 
 \textbf{rs7219773} & TNK1 & Parahipp(R)\\ 

\end{longtable}
\end{center}
\normalsize

\pagebreak

\section*{Web Appendix B: Derivations for the Gibbs Sampling and Mean Field Variational Bayes Algorithm}
Recall that $\vomega^{2}=(\omegao, \cdots, \omegad)$, $\tilde {W}_{i,j^{*}}=({W}_{i,j},W_{i,{j+1}})$,~$i=1,...,d,~j=2j^{*}-1$,~~$j^{*}=1,...,\frac{c}{2}$. based on the hierarchical prior setting, the joint posterior distribution can be expressed up to a normalizing constant as
\small
\begin{equation*}
\begin{split}
     p  (\boldsymbol{W},\omegao, \cdots, \omegad, \boldsymbol{{\boldsymbol{\Sigma}}},| \boldsymbol{Y} ) {}  &\propto{}  p(\boldsymbol{Y} | \boldsymbol{W}, \boldsymbol{{\boldsymbol{\Sigma}}}) p(\boldsymbol{W} | \boldsymbol{{\boldsymbol{\Sigma}}}, \vomega^{2})p(\vomega^{2} )p(\boldsymbol{{\boldsymbol{\Sigma}}} )\\
&\propto{}  |(D_A-\rho A)^{-1}\otimes\boldsymbol{{\boldsymbol{\Sigma}}}|^{-\frac{n}{2}} \exp \Bigg\lbrace - \frac{1}{2}\sum_{\ell=1}^n ( \boldsymbol{y}_{\ell} - \boldsymbol{W}^T\boldsymbol{x}_{\ell})^T [(D_A-\rho A)^{-1}\otimes\boldsymbol{{\boldsymbol{\Sigma}}}]^{-1}\\
&(\boldsymbol{y}_{\ell} - \boldsymbol{W}^T\boldsymbol{x}_{\ell}) \Bigg \rbrace \\
& \times \prodid (\omega_{i}^2)^{-\frac{c}{2}}|\boldsymbol{{\boldsymbol{\Sigma}}}|^{-\frac{c}{4}}\exp\{-\frac{1}{2}\sum_{j=1}^{\frac{c}{2}}\tilde{W}_{i,j}^T(\omega_i^2\boldsymbol{{\boldsymbol{\Sigma}}})^{-1}\tilde{W}_{i,j}\} \\
&\times \prodid \frac{{(\frac{\lambda^2}{2})^{\frac{c+1}{2}}}}{\Gamma(\frac{c+1}{2})}(\omega_i^2)^{\frac{c}{2}-\frac{1}{2}} \exp\{-\frac{\lambda^2}{2}\omega_i^2\}\\
&\times \frac{|S|^{\frac{v}{2}}}{2^{v}\Gamma _2(\frac{v}{2})}|\boldsymbol{{\boldsymbol{\Sigma}}}|^{-\frac{v+3}{2}} \exp \left\lbrace - \frac{1}{2} tr(S\boldsymbol{{\boldsymbol{\Sigma}}}^{-1}) \right\rbrace \\
\end{split}
\end{equation*}
\normalsize
%
\textbf{\emph{The full conditional distribution of }$\boldsymbol{W_{(i)}}$}

{For any matrix $A$, let $A_{(i)}$ be the $i_{th}$ row of  $A$, $A_{i,j}$ be  $i_{th}$ row and $j_{th}$ column of $A$ if $A$ is a matrix and  $A_{(i)}$ be the $i_{th}$ element of  $A$ if $A$ is a vector. And let $A_{(-i)}$ be the rest after removing   $A_{(i)}$ from $A$.} The full conditional distribution of $\boldsymbol{W_{(i)}}, \; i=1, \dots, d$ is expressed as 
\small
$$ (\boldsymbol{W}_{(i)})^{T} \big| \boldsymbol{Y}, \boldsymbol{W}_{(-i)}, \vomega, \boldsymbol{{\boldsymbol{\Sigma}}} \sim MVN_{c} ( \; \bvmui ,\; \boldsymbol{{\boldsymbol{\Sigma}}}_i ),$$ 
\normalsize
where 
\small
\begin{align*}
\bvmui &=\boldsymbol{{\boldsymbol{\Sigma}}}_i \Bigg (- \sumln (\boldsymbol{x}_{\ell (i)} \otimes I_c)[(D_A-\rho A)\otimes\boldsymbol{{\boldsymbol{\Sigma}}}^{-1}](\boldsymbol{x}_{\ell (-i)}^T \otimes I_c) (\boldsymbol{W}_{(-i)}^T) \\
&+ \sumln (\boldsymbol{x}_{\ell (i)} \otimes I_c)[(D_A-\rho A)\otimes\boldsymbol{{\boldsymbol{\Sigma}}}^{-1}] \boldsymbol{y}_{\ell} \Bigg)  
\end{align*}
\normalsize
\small
$$\boldsymbol{{\boldsymbol{\Sigma}}}_i = \left(\boldsymbol{H}_i+\sumln (\boldsymbol{x}_{\ell (i)} \otimes I_c)[(D_A-\rho A)\otimes\boldsymbol{{\boldsymbol{\Sigma}}}^{-1}](\boldsymbol{x}_{\ell (i)}^{T} \otimes I_c)\right)^{-1}.$$
$$ \boldsymbol{H}_i = \left[ \left\lbrace \frac{1}{\omega_i^2} \right\rbrace\otimes I_{\frac{c}{2}}\otimes \boldsymbol{{\boldsymbol{\Sigma}}}^{-1}\right].$$
\normalsize
Since we already have the full conditional distribution, the coordinate-wise updates for mean field variational Bayes can be derived as:
\small
\begin{align*}
q((\boldsymbol{W}_{(i)})^{T}) & \propto  exp \Big\{{\bf{E}}_{-i}(log P((\boldsymbol{W}_{(i)})^{T}|rest))\Big\}\\
&\propto exp \Big\{{\bf{E}}_{-i}\Big( -\frac{c}{2} log(2\pi) -\frac{1}{2} log(det|\boldsymbol{{\boldsymbol{\Sigma}}}_{i}|) -  \frac{1}{2} ((\boldsymbol{W}_{(i)})^{T} - \bvmui)^{T} \boldsymbol{{\boldsymbol{\Sigma}}}_i ^{-1}((\boldsymbol{W}_{(i)})^{T} - \bvmui)\Big)\Big\} \\
&\propto exp \Big\{{\bf{E}}_{-i}\Big( - const  -  \frac{1}{2} (\boldsymbol{W}_{(i)})\boldsymbol{{\boldsymbol{\Sigma}}}_i ^{-1} (\boldsymbol{W}_{(i)})^{T} +(\boldsymbol{W}_{(i)}) \boldsymbol{{\boldsymbol{\Sigma}}}_i ^{-1} \bvmui\Big)\Big\} \\
&\propto exp\Big\{   const -  \frac{1}{2} (\boldsymbol{W}_{(i)}){\bf{E}}_{-i}(\boldsymbol{{\boldsymbol{\Sigma}}}_i ^{-1} )(\boldsymbol{W}_{(i)})^{T}  + (\boldsymbol{W}_{(i)}) {\bf E}_{-i} \big(\boldsymbol{{\boldsymbol{\Sigma}}}_i ^{-1} \bvmui \big) \Big\}
\end{align*}
\normalsize

We still can see that $q((\boldsymbol{W}_{(i)})^{T})$ is still MVN with 
\small
\begin{align*}
 \vSigma_{q{(\boldsymbol{W}_{(i)})}}^{-1} &= {\bf{E}}_{-i}(\boldsymbol{{\boldsymbol{\Sigma}}}_i ^{-1} )\\
 & =  E_{-i}\left(\left[  \left\lbrace \frac{1}{\omega_i^2} \right\rbrace\otimes I_{\frac{c}{2}}\otimes \boldsymbol{{\boldsymbol{\Sigma}}}^{-1}\right]+\sumln (\boldsymbol{x}_{\ell (i)} \otimes I_c)[(D_A-\rho A)\otimes\boldsymbol{{\boldsymbol{\Sigma}}}^{-1}](\boldsymbol{x}_{\ell (i)}^{T} \otimes I_c)\right).\\
 & = \bigg (  \left[ \left\lbrace \frac{1}{\omega_i^2} \right\rbrace\otimes I_{\frac{c}{2}}\otimes E_{q(\boldsymbol{{\boldsymbol{\Sigma}}})}(\boldsymbol{{\boldsymbol{\Sigma}}}^{-1}) \right]  +\sumln (\boldsymbol{x}_{\ell (i)} \otimes I_c)[(D_A- \rho A)\otimes E_{q(\boldsymbol{{\boldsymbol{\Sigma}}})}(\boldsymbol{{\boldsymbol{\Sigma}}}^{-1})](\boldsymbol{x}_{\ell (i)}^{T} \otimes I_c) \bigg )
\end{align*}
\normalsize
Also, we can find that:
\small
\begin{align*}
  {\bf E}_{-i} \big(\boldsymbol{{\boldsymbol{\Sigma}}}_i ^{-1} \bvmui \big)  &= \vSigma_{q{(\boldsymbol{W}_{(i)})}}^{-1}\vmu_{q(\boldsymbol{W}_{(i)})}  = {\bf E}_{-i} \bigg(- \sumln (\boldsymbol{x}_{\ell (i)} \otimes I_c)[(D_A-\rho A)\otimes\boldsymbol{{\boldsymbol{\Sigma}}}^{-1}](\boldsymbol{x}_{\ell (-i)}^T \otimes I_c) (\boldsymbol{W}_{(-i)}^T) \\
  &+ \sumln (\boldsymbol{x}_{\ell (i)} \otimes I_c)[(D_A-\rho A)\otimes\boldsymbol{{\boldsymbol{\Sigma}}}^{-1}] \boldsymbol{y}_{\ell} \bigg)\\
         \Rightarrow \vSigma_{q{(\boldsymbol{W}_{(i)})}}^{-1}\vmu_{q(\boldsymbol{W}_{(i)})} &=   \bigg(- \sumln (\boldsymbol{x}_{\ell (i)} \otimes I_c)[(D_A- \rho  A)\otimes E_{q({\boldsymbol{\Sigma}})}({\boldsymbol{\Sigma}}^{-1})](\boldsymbol{x}_{\ell (-i)}^T \otimes I_c) \big(\vmu_{q(\boldsymbol{W}_{(-i)})}\big) \\
         &+ \sumln (\boldsymbol{x}_{\ell (i)} \otimes I_c)[(D_A- \rho  A)\otimes E_{q({\boldsymbol{\Sigma}})}({\boldsymbol{\Sigma}}^{-1})] \boldsymbol{y}_{\ell} \bigg)   \\
         \Rightarrow \vmu_{q(\boldsymbol{W}_{(i)})} & = \vSigma_{q{(\boldsymbol{W}_{(i)})}} \bigg(- \sumln (\boldsymbol{x}_{\ell (i)} \otimes I_c)[(D_A- \rho  A)\otimes E_{q({\boldsymbol{\Sigma}})}({\boldsymbol{\Sigma}}^{-1})](\boldsymbol{x}_{\ell (-i)}^T \otimes I_c) \big(\vmu_{q(\boldsymbol{W}_{(-i)})}\big) \\
         &+ \sumln (\boldsymbol{x}_{\ell (i)} \otimes I_c)[(D_A- \rho  A)\otimes E_{q({\boldsymbol{\Sigma}})}({\boldsymbol{\Sigma}}^{-1})] \boldsymbol{y}_{\ell} \bigg) 
\end{align*}
\normalsize
Then, we can also compute:
	\small
	\begin{align*}
	 E_q [log(q(vec \vW_{(i)}^{T}))] &= E_q\Bigg[ -\frac{1}{2} log|2 \pi {\boldsymbol{\Sigma}}_{q{(\boldsymbol{W}_{(i)})}}| - \frac{1}{2}\left (vec (\vW_{(i)}^{T} )- \vmu_{q{(\boldsymbol{W}_{(i)})}} \right)^{T} {\boldsymbol{\Sigma}}_{q{(\boldsymbol{W}_{(i)})}} ^{-1}\\
	 &\left (vec( \vW_{(i)}^{T}) - \vmu_{q{(\boldsymbol{W}_{(i)})}} \right)\Bigg] \\
	 & =   -\frac{1}{2} log|2 \pi {\boldsymbol{\Sigma}}_{q{(\boldsymbol{W}_{(i)})}}| - \frac{1}{ 2} \vmu_{q{(\boldsymbol{W}_{(i)})}}^T {\boldsymbol{\Sigma}}_{q{(\boldsymbol{W}_{(i)})}} ^{-1}\vmu_{q{(\boldsymbol{W}_{(i)})}}  \\
	 &+ E_q\bigg[ -  \frac{1}{2} (\boldsymbol{W}_{(i)})\boldsymbol{{\boldsymbol{\Sigma}}}_{q(\boldsymbol{W}_{(i)})} ^{-1} (\boldsymbol{W}_{(i)})^{T}  +(\boldsymbol{W}_{(i)}) \boldsymbol{{\boldsymbol{\Sigma}}}_{q(\boldsymbol{W}_{(i)})} ^{-1} \vmu_{q(\boldsymbol{W}_{(i)})}\ \bigg ] \\
	 & =   -\frac{1}{2} log|2 \pi {\boldsymbol{\Sigma}}_{q{(\boldsymbol{W}_{(i)})}}| - \frac{1}{ 2} \vmu_{q{(\boldsymbol{W}_{(i)})}}^T {\boldsymbol{\Sigma}}_{q{(\boldsymbol{W}_{(i)})}} ^{-1}\vmu_{q{(\boldsymbol{W}_{(i)})}} \\
	 & -\frac{1}{2} \Big(\vmu_{q{(\boldsymbol{W}_{(i)})}} ^T \boldsymbol{{\boldsymbol{\Sigma}}}_{q(\boldsymbol{W}_{(i)})} ^{-1} \vmu_{q{(\boldsymbol{W}_{(i)})}}  +  tr(\boldsymbol{{\boldsymbol{\Sigma}}}_{q{(\boldsymbol{W}_{(i)})}} ^{-1}\boldsymbol{{\boldsymbol{\Sigma}}}_{q{(\boldsymbol{W}_{(i)})}} )\Big) + \vmu_{q{(\boldsymbol{W}_{(i)})}} ^T\boldsymbol{{\boldsymbol{\Sigma}}}_{q(\boldsymbol{W}_{(i)})} ^{-1} \vmu_{q(\boldsymbol{W}_{(i)})}\\
	 & = -\frac{1}{2} log|2 \pi {\boldsymbol{\Sigma}}_{q{(\boldsymbol{W}_{(i)})}}| - \frac{1}{ 2}tr(\boldsymbol{{\boldsymbol{\Sigma}}}_{q{(\boldsymbol{W}_{(i)})}} ^{-1}\boldsymbol{{\boldsymbol{\Sigma}}}_{q{(\boldsymbol{W}_{(i)})}} )
	\end{align*}
	\normalsize

\textbf{\noindent{\emph{Full conditional distribution of }${\boldsymbol{\Sigma}}$}:}
\small
\begin{align*}
\begin{split}
p( {\boldsymbol{\Sigma}} \big|  \boldsymbol{Y}, \boldsymbol{W},\vomega^2)  &\propto{}  p(\boldsymbol{Y} | \boldsymbol{W}, {\boldsymbol{\Sigma}}) p(\boldsymbol{W} | {\boldsymbol{\Sigma}}, \vomega^{2})p({\boldsymbol{\Sigma}} )\\
&\propto{}|(D_A-\rho A)^{-1}\otimes{\boldsymbol{\Sigma}}|^{-\frac{n}{2}} \exp \Bigg\lbrace - \frac{1}{2}\sum_{\ell=1}^n ( \boldsymbol{y}_{\ell} - \boldsymbol{W}^T\boldsymbol{x}_{\ell})^T [(D_A-\rho A)\otimes{\boldsymbol{\Sigma}} ^{-1}]\\
&(\boldsymbol{y}_{\ell} - \boldsymbol{W}^T\boldsymbol{x}_{\ell}) \Bigg\rbrace \\ 
& \times  \prodid {|\omega_i^2{\boldsymbol{\Sigma}}|}^{-\frac{c}{4}}  \exp \left\lbrace - \frac{1}{2}{ \sum\limits_{j^{*}=1}^{\frac{c}{2}} {\tilde{W}_{i,j^{*}}}^T ~(\omega_i^2{\boldsymbol{\Sigma}})^{-1}~ {\tilde{W}_{i,j^{*}}}}  \right\rbrace  \\
&\times \frac{|S|^{\frac{v}{2}}}{2^{v}\Gamma _2(\frac{v}{2})}|{\boldsymbol{\Sigma}}|^{-\frac{v+3}{2}} \exp \left\lbrace - \frac{1}{2} tr(S{\boldsymbol{\Sigma}}^{-1}) \right\rbrace \\ 
\end{split}
\end{align*}
\normalsize
Denote~~$y^*_l=y_l-\vW^Tx_l$,~~ ${\tilde{y^*}_{l,j^*}}^{T}=(y^*_{l,j},y^*_{l,j+1})$,~~$j=2j^{*}-1 $,~~$j^{*}=1,...,\frac{c}{2}$,~~$l=1,...,n$. Let 
$B=D_A-\rho A$,~~ then dim($B$)=$\frac{c}{2}\times \frac{c}{2}$.~~  Denote $b_{i,j}$ be  $i_{th}$~row and $j_{th}$ column of $B$,~~$b_{i,j}$ is a scalar, where $i=1,...,\frac{c}{2},j=1,...,\frac{c}{2}$.
Using $|E \otimes F| = |E|^n|F|^m$, where $dim(E)=n \times n$ and $dim(F)= m \times m$. 
$tr(G)+tr(Q)=tr(G+Q)$ where dim(G)=dim(Q) and 
$tr(JK)=tr(KJ)$ where dim($J$)=dim($K^T$). This can be simplified as:
\small
\begin{align*}
\begin{split}
p( {\boldsymbol{\Sigma}} \big|  \boldsymbol{Y}, \boldsymbol{W},\vomega^2) \propto{}&|D_A-\rho A|^{\frac{nc}{4}}|{\boldsymbol{\Sigma}}|^{-n} \exp \left\lbrace - \frac{1}{2} tr\left(\sum\limits_{l=1}^{n} \sum\limits_{i=1}^{\frac{c}{2}} \sum\limits_{j=1}^{\frac{c}{2}}b_{i,j}\tilde{y^*_{l,i}}{\tilde{y^*_{l,i}}}^T {\boldsymbol{\Sigma}}^{-1}\right) \right\rbrace \\ 
& \times  \prodid {|\omega_i^2{\boldsymbol{\Sigma}}|}^{-\frac{c}{4}}  \exp \left\lbrace - \frac{1}{2}{ \sum\limits_{j^{*}=1}^{\frac{c}{2}} {\tilde{W}_{i,j^{*}}}^T ~(\omega_i^2{\boldsymbol{\Sigma}})^{-1}~ {\tilde{W}_{i,j^{*}}}}  \right\rbrace  \\
&\times \frac{|S|^{\frac{v}{2}}}{2^{v}\Gamma _2(\frac{v}{2})}|{\boldsymbol{\Sigma}}|^{-\frac{v+3}{2}} \exp \left\lbrace - \frac{1}{2} tr(S{\boldsymbol{\Sigma}}^{-1})  \right\rbrace.
\end{split}
\end{align*}
\normalsize
Since~~ $|D_A-\rho A|$~~,~~{} $\prodid {|\omega_i^2|}^{-\frac{c}{2}} $ and  $\frac{|S|^{\frac{v}{2}}}{2^{v}\Gamma _2(\frac{v}{2})}$ do not depend on ${\boldsymbol{\Sigma}}$, they can be factored out of the expression. This leaves, 
\small
\begin{align*}
\begin{split}
p( {\boldsymbol{\Sigma}} \big|  \boldsymbol{Y}, \boldsymbol{W},\vomega^2) &\propto{}|{\boldsymbol{\Sigma}}|^{-n}  \exp \left\lbrace - \frac{1}{2} tr\left(\sum\limits_{l=1}^{n} \sum\limits_{i=1}^{\frac{c}{2}} \sum\limits_{j=1}^{\frac{c}{2}}b_{i,j}\tilde{y^*_{l,i}}{\tilde{y^*_{l,i}}}^T {\boldsymbol{\Sigma}}^{-1}\right)\right\rbrace \\ 
& \times |{\boldsymbol{\Sigma}}|^{-\frac{cd}{4}} \exp \left\lbrace - \frac{1}{2}tr\left(\sum\limits_{i=1}^{d}\sum\limits_{j^{*}=1}^{\frac{c}{2}}\frac{  \tilde{W_{i,j^{*}}} {\tilde{W_{i,j^{*}}}}^T}{\omega_i^2} {\boldsymbol{\Sigma}}^{-1}\right) \right\rbrace  \\
&\times |{\boldsymbol{\Sigma}}|^{-\frac{v+3}{2}} \exp \left\lbrace - \frac{1}{2} tr(S{\boldsymbol{\Sigma}}^{-1})  \right\rbrace.\\
\end{split}
\end{align*}
\begin{align*}
\begin{split}
\propto{} |{\boldsymbol{\Sigma}}|^{-\frac{2n+\frac{cd}{2}+v+3}{2}}  \exp \left\lbrace - \frac{1}{2} tr\left[\left(\sum\limits_{l=1}^{n} \sum\limits_{i=1}^{\frac{c}{2}} \sum\limits_{j=1}^{\frac{c}{2}}b_{i,j}\tilde{y^*_{l,i}}{\tilde{y^*_{l,i}}}^T +
\sum\limits_{i=1}^{d}\sum\limits_{j^{*}=1}^{\frac{c}{2}}\frac{  \tilde{W}_{i,j^{*}} {\tilde{W}_{i,j^{*}}}^T}{\omega_i^2} +S
\right){\boldsymbol{\Sigma}}^{-1} \right] \right\rbrace
\end{split}
\end{align*}

{ \normalsize Therefore}\\
\begin{align*}
\begin{split}
{\boldsymbol{\Sigma}} \sim Inverse-Wishart(S^*,v^*)
\end{split}
\end{align*}

{\normalsize Where}\\
$$S^*=\sum\limits_{l=1}^{n} \sum\limits_{i=1}^{\frac{c}{2}} \sum\limits_{j=1}^{\frac{c}{2}}b_{i,j}\tilde{y^*_{l,i}}{\tilde{y^*_{l,i}}}^T +\sum\limits_{i=1}^{d}\sum\limits_{j^{*}=1}^{\frac{c}{2}}\frac{  \tilde{W}_{i,j^{*}} {\tilde{W}_{i,j^{*}}}^T}{\omega_i^2}+S$$

 $$v^*=2n+\frac{cd}{2}+v$$ 
 
 \normalsize
 $S^*$ is a $2\times2$ matrix and $v^*$ is a scalar. Similarly, we can derive the  mean field approximation for ${\boldsymbol{\Sigma}}$  based on the full conditional distribution as follows:
\small
\begin{align*}
q({\boldsymbol{\Sigma}}) &\propto exp\Big\{ {\bf E}_{rest} \Big( log P({\boldsymbol{\Sigma}} |rest)\Big) \Big\}\\
&\propto exp\Big\{ {\bf E}_{rest} \Big(-\frac{2n+\frac{cd}{2}+v+3}{2} log(|{\boldsymbol{\Sigma}}| ) - \frac{1}{2} tr\Bigg[\Bigg(\sum\limits_{l=1}^{n} \sum\limits_{i=1}^{\frac{c}{2}} \sum\limits_{j=1}^{\frac{c}{2}}b_{i,j}\tilde{y^*_{l,i}}{\tilde{y^*_{l,i}}}^T \\
& +\sum\limits_{i=1}^{d}\sum\limits_{j^{*}=1}^{\frac{c}{2}}\frac{  \tilde{W_{i,j^{*}}} {\tilde{W_{i,j^{*}}}}^T}{\omega_i^2} +S
\Bigg){\boldsymbol{\Sigma}}^{-1} \Bigg] 
 \Big) \Big\}\\
 &\propto exp\Bigg\{ -\frac{2n+\frac{cd}{2}+v+3}{2} log(|{\boldsymbol{\Sigma}}| ) - {\bf E}_{rest} \Big(\frac{1}{2} tr\Bigg[\bigg(\sum\limits_{l=1}^{n} \sum\limits_{i=1}^{\frac{c}{2}} \sum\limits_{j=1}^{\frac{c}{2}}b_{i,j}\tilde{y^*_{l,i}}{\tilde{y^*_{l,i}}}^T \\
 & +\sum\limits_{i=1}^{d}\sum\limits_{j^{*}=1}^{\frac{c}{2}}\frac{  \tilde{W_{i,j^{*}}} {\tilde{W_{i,j^{*}}}}^T}{\omega_i^2} +S
\bigg){\boldsymbol{\Sigma}}^{-1} \Bigg] 
 \Big) \Bigg\}
\end{align*}
\normalsize
This is still a Inverse-Wishart distribution.  That is
\begin{align*}
q({\boldsymbol{\Sigma}})  \sim Inverse-Wishart(S_{q({\boldsymbol{\Sigma}})}, v_{q({\boldsymbol{\Sigma}})})
\end{align*}
 For $\tilde {W}_{i,j^{*}}=({W}_{i,j},W_{i,{j+1}})$,~~$j=2j^{*}-1$,~~$j^{*}=1,...,\frac{c}{2}$. 

%
\small
\[
E_{q(\boldsymbol{W})}(\tilde{W}_{i,j^{*}}{\tilde{W}_{i,j^{*}}}^T) =
\begin{bmatrix}
  E_{q(\boldsymbol{W})}({W_{i,j}^2}) & E_{q(\boldsymbol{W})}({W_{i,j}W_{i,j+1}})\\
  E_{q(\boldsymbol{W})}({W_{i,j}W_{i,j+1}}) & E_{q(\boldsymbol{W})}({W_{i,j+1}^2})
   \end{bmatrix}
\]
\normalsize
Now, for  $E_{q(\boldsymbol{W})}({W_{i,j}^2})$ and $E_{q(\boldsymbol{W})}({W_{i,j}W_{i,j+1}})$, we can get that:
	\small
	\begin{align*}
	E_{q(\boldsymbol{W})}({W_{i,j}^2}) & =  ({\vmu_{q(\boldsymbol{W}_{(i)})}}_{(j)})^2 + {\vSigma_{q(\boldsymbol{W}_{(i)})}}_{j,j}\\
	E_{q(\boldsymbol{W})}({W_{i,j+1}^2}) & =  ({\vmu_{q(\boldsymbol{W}_{(i)})}}_{(j+1)})^2 + {\vSigma_{q(\boldsymbol{W}_{(i)})}}_{{j+1,j+1}}\\
	E_{q(\boldsymbol{W})}({W_{i,j}W_{i,j+1}}) & = {\vmu_{q(\boldsymbol{W}_{(i)})}}_{(j)}  {\vmu_{q(\boldsymbol{W}_{(i)})}}_{(j+1)}+{\vSigma_{q(\boldsymbol{W}_{(i)})}}_{j,j+1}
	\end{align*}
\begin{align*}
S_{q({\boldsymbol{\Sigma}})} &= E_{rest}\left(\sum\limits_{l=1}^{n} \sum\limits_{i=1}^{\frac{c}{2}} \sum\limits_{j=1}^{\frac{c}{2}}b_{i,j}\tilde{y^*_{l,i}}{\tilde{y^*_{l,i}}}^T +
\sum\limits_{i=1}^{d}\sum\limits_{j^{*}=1}^{\frac{c}{2}}\frac{  \tilde{W_{i,j^{*}}} {\tilde{W_{i,j^{*}}}}^T}{\omega_i^2} +S
\right) \\
& = \left(\sum\limits_{l=1}^{n} \sum\limits_{i=1}^{\frac{c}{2}} \sum\limits_{j=1}^{\frac{c}{2}} b_{i,j}\tilde{y^*_{l,i}}{\tilde{y^*_{l,i}}}^T +
\sum\limits_{i=1}^{d}\sum\limits_{j^{*}=1}^{\frac{c}{2}}{ E_q\Big( \tilde{W_{i,j^{*}}} {\tilde{W_{i,j^{*}}}}^T\Big)}E_q\Big(\frac{1}{\omega_i^2}\Big) +S
\right) 
\end{align*}
$$v_{q({\boldsymbol{\Sigma}})} = 2n+\frac{cd}{2}+v $$

Now for $log(q({\boldsymbol{\Sigma}}))$:
	\begin{align*}
	E_q(log(q({\boldsymbol{\Sigma}}))) & = E_q \left[ \frac{v_{q({\boldsymbol{\Sigma}})}}{2} log|S_{q({\boldsymbol{\Sigma}})}| - log(2^{v_{q({\boldsymbol{\Sigma}})}}) - log\Gamma _2(\frac{v_{q({\boldsymbol{\Sigma}})}}{2}) -\frac{v+3}{2} log(|{\boldsymbol{\Sigma}}|)   - \frac{1}{2} tr(S_{q({\boldsymbol{\Sigma}})}{\boldsymbol{\Sigma}}^{-1}) \right] \\
	& =\frac{v_{q({\boldsymbol{\Sigma}})}}{2} log|S_{q({\boldsymbol{\Sigma}})}| - log(2^{v_{q({\boldsymbol{\Sigma}})}}) - log\Gamma _2(\frac{v_{q({\boldsymbol{\Sigma}})}}{2}) -  E_q \left[ \frac{v+3}{2} log(|{\boldsymbol{\Sigma}}|)   - \frac{1}{2} tr(S_{q({\boldsymbol{\Sigma}})}{\boldsymbol{\Sigma}}^{-1}) \right]
	\end{align*}
	\normalsize

\textbf{\noindent{\emph{Full Conditional of $\vomega^{2}$}}:}

We consider a joint update of the scale mixing variable based on the corresponding full conditional distribution. We have:
\small
\begin{align*}
\begin{split}
p(\vomega^{2} \big| \boldsymbol{Y}, \boldsymbol{W}, {\boldsymbol{\Sigma}})  &\propto{} p(\boldsymbol{W} | {\boldsymbol{\Sigma}}, \vomega^{2})p(\vomega^{2} |\lambda^2 )\\
&\propto{}  \prodid (\omega_i^2)^{-\frac{c}{2}}{|{\boldsymbol{\Sigma}}|}^{-\frac{c}{4}}  \exp \left\lbrace - \frac{1}{2}{ \sum\limits_{j^{*}=1}^{\frac{c}{2}} {\tilde{w}_{i,j^{*}}}^T ~(\omega_i^2{\boldsymbol{\Sigma}})^{-1}~ {\tilde{w}_{i,j^{*}}}}  \right\rbrace\\
&\times \prodid \frac{{(\frac{\lambda^2}{2})^{\frac{c+1}{2}}}}{\Gamma(\frac{c+1}{2})}(\omega_i^2)^{\frac{c+1}{2}-1} \exp\{-\frac{\lambda^2}{2}\omega_i^2\}
\end{split}
\end{align*}
\begin{align*}
\begin{split}
\propto  \prodid (\omega_i^2)^{-\frac{1}{2}} \exp \left\lbrace -\left(\frac{\lambda^2}{2}\right)\omega_{i}^{2} - \frac{c_i^*}{2\omega_{i}^{2} } \right\rbrace
\end{split}
\end{align*}

{\normalsize where}: 
$$c_i^{*}=  \sum\limits_{j^{*}=1}^{\frac{c}{2}}
{\tilde{W}_{i,j^{*}}}^T{\boldsymbol{\Sigma}}^{-1}
{\tilde{W}_{i,j^{*}}}=tr(  \sum\limits_{j^{*}=1}^{\frac{c}{2}}{  \tilde{W}_{i,j^{*}} {\tilde{W}_{i,j^{*}}}^T{\boldsymbol{\Sigma}}^{-1}})$$ 
\normalsize

The above expression shows that the scale mixing variables are conditionally independent given $\boldsymbol{Y}, \boldsymbol{W}, {\boldsymbol{\Sigma}},\rho, \lambda^2$. We next apply a transformation of variable  $\eta_{i}  = (\omega_{i}^{2})^{-1}$, Jacobian = $\big| \frac{d}{d \eta_{i}} \omega_{i}^{2} \big| =  \eta_{i}^{-2}$ which yields:
\small
\begin{align*}
\begin{split}
p(\vomega^{2} \big| \boldsymbol{Y}, \boldsymbol{W}, {\boldsymbol{\Sigma}}, \rho,\lambda^2) \propto  &  \prodid (\eta_i)^{-\frac{3}{2}} \exp \left\lbrace -\left(\frac{\lambda^2}{2\eta_{i}}\right) - \frac{\eta_{i}c_{i}^{*}}{2} \right\rbrace
\end{split}
\end{align*}
\normalsize
and from this we see that the conditional distributions lie within the Inverse Gaussian family. More specifically we have
$$\eta_i= \frac{1}{\omegai} \; \;\Big| \; \boldsymbol{Y}, \boldsymbol{W}, {\boldsymbol{\Sigma}},\rho, \lambda^2  \sim \textit{Inverse-Gaussian} \left( \sqrt{ \frac{\lambda^2}{c_i^*}} \;, \;\;\lambda^2 \right), \,\,\, i=1,\dots,d.$$
Now, since we already know the full conditional distribution of $\omega_i$, we have:
\small
\begin{align*}
q(\eta_i) &\propto exp\Big\{ {\bf E}_{q} \Big( log P(\eta_i |rest)\Big) \Big\}\\
 &\propto exp\Big\{ {\bf E}_{q} \Big( -\frac{3}{2}log(\eta_i)  -\left(\frac{\lambda^2}{2\omega_{i}^2}\right) - \frac{\omega_{i}^2c_{i}^{*}}{2} \Big) \Big\}\\
 &\propto exp\left\{\Big( -\frac{3}{2}log(\eta_i)  -  \left( \frac{\lambda^2}{2\eta_i}\right) - {\bf E}_{q} \left(\frac{\eta_ic_{i}^{*}}{2}\right) \Big) \right\}\\
 &\propto exp\left\{\Big( -\frac{3}{2}log(\eta_i)  -  \left( \frac{\lambda^2}{2}\right) \frac{1}{\eta_i} - {\bf E}_{c_i^*}(c_{i}^{*}) \left(\frac{\eta_i}{2}\right) \Big) \right\}\\
\end{align*}
\normalsize
Therefore, $q(\eta_i) $ is still an Inverse-Gaussian distribution with
\small
$$
\mu_{q(\eta_i)} = \sqrt{\frac{{\lambda^2}}{{ E}_{c_i^*}(c_{i}^{*}) }}  $$,
$$E_{c_i^*}(c_{i}^{*})=\sum\limits_{j^{*}=1}^{\frac{c}{2}} E(tr(  {  \tilde{W}_{i,j^{*}} {\tilde{W}_{i,j^{*}}}^T{\boldsymbol{\Sigma}}^{-1}}))=\sum\limits_{j^{*}=1}^{\frac{c}{2}} (E_{q(\boldsymbol{W})}(W_{i,j}^2)+E_{q(\boldsymbol{W})}(W_{i,j+1}^2))tr(S_{q(\boldsymbol{\Sigma})}^{-1})$$

\normalsize
		
Now, since $\eta_i$ is an Inverse Gaussian, then $\omega_i^2 =  1 / \eta_i$ will be a reciprocal of inverse gaussian.  where we have:
	\small
	\begin{align*}
	\mu_{q(\omega_i^2) }& =  \frac{1}{\mu_{q(\eta_i)}} + \frac{1}{\lambda^2 }\\
	Var_{q(\omega_i^2)} &  = \frac{1}{\mu_{q(\eta_i)}\lambda^2} + \frac{2}{{(\lambda^2)} ^2}
	\end{align*}

For $E_q(log(q(\omega_i^2)))$, we can compute as:
	\begin{align*}
	E_q(log(q(\omega_i^2))) & = E_q\left[ \frac{1}{2} \left( log({\lambda^2} ) -  log(2\pi) - log(\omega_i^2)\right)  - \frac{\lambda^2 (1 - \omega_i^2\mu_{q(\eta_i)})^2}{2\mu_{q(\eta_i)}^2 \omega_i^2}\right] \\
	& =  \frac{1}{2} \left( log({\lambda^2} ) -  log(2\pi) \right)  -  E_q\left[  log(\omega_i^2) \right] -  {\lambda^2}  E_q\left[  \frac{1}{2\mu_{q_{(\eta_i)}}^2}\left( \frac{1}{\omega_i^2}\right) - \frac{1}{\mu_{q_{(\eta_i)}}^2}  + \frac{\omega_i^2}{2}\right]
	\end{align*}
For $E_q(log(\omega_i^2))$, we can use Taylor series to approximate as:
	$$
	E_q(log(\omega_i^2)) =  log(\mu_{q(\omega_i^2)}) - \frac{1}{2\mu_{\omega_i^2}}Var_q[\omega_i^2]
$$
\normalsize

Then, we can have:
\small
	\begin{align*}
	E_q(log(q(\omega_i^2))) & =  \frac{1}{2} \left( log({\lambda^2} ) -  log(2\pi) \right)  - log(\mu_{q(\omega_i^2})) - \frac{1}{2\mu_{\omega_i^2}}Var_q[\omega_i^2] \\
	& -{\lambda^2}  E_q\left[  \frac{1}{2\mu_{q_{(\eta_i)}}^2}\left( \frac{1}{\omega_i^2}\right) - \frac{1}{\mu_{q_{(\eta_i})}^2}  + \frac{\omega_i^2}{2}\right]\\
	& =  \frac{1}{2} \left( log({\lambda^2} ) -  log(2\pi) \right)  - log(\mu_{q(\omega_i^2)}) - \frac{1}{2\mu_{\omega_i^2}}Var_q[\omega_i^2] \\
	&-{\lambda^2}  \left[\frac{1}{2\mu_{q_{(\eta_i)}}^2}  \mu_{q(\eta_i)} - \frac{1}{\mu_{q_{(\eta_i)}}^2}  + \frac{\mu_{q(\omega_i^2)}}{2}\right]
	\end{align*}
\normalsize
Therefore, the posterior distribution can be approximated by mean field variational bayes as:
	\small
	\begin{align*}
	P(\vTheta | Y) \approx \prod_{i=1}^{d} \Big[ q_{\vW_{(i)}}(\vW_{(i)}) \Big]  \prod_{i = 1}^{d}\Big[q_{\omega^2_i}(\omega^2_i)\Big]q_{{\boldsymbol{\Sigma}}}({\boldsymbol{\Sigma}})
	\end{align*}
	\normalsize
where:
	\small
	\begin{align*}
	q_{\vW_{(i)}}(\vW_{(i)}) &\equiv MVN(\vmu_{q_{\vW_{(i)}}},  \vSigma_{q_{\vW_{(i)}}})\\
	q_{\omega^2_i}(\omega^2_i) &\equiv \text{reciprocal of Inverse Gaussian}(\mu_{q(\eta_i)} , \lambda^2 )\\
	q_{{\boldsymbol{\Sigma}}}({\boldsymbol{\Sigma}}) &\equiv  Inverse-Wishart(S_{q({\boldsymbol{\Sigma}})}, v_{q({\boldsymbol{\Sigma}})})
\\
\end{align*}
\normalsize
\textbf{\emph{Lower Bound $\mathcal{L}(q) $ for Variational Bayes}.}

We now have derived the optimal $q$ distributions. The logarithm lower bound takes following explicit  form:
	\small
	\begin{align*}
	\mathcal{L}(q) & =  E_{q}[log(P(\vY, \vtheta))] - E_{q}[log(q(\vtheta))]\\
	& = E_{q} \big[ log(p(\boldsymbol{Y} | \boldsymbol{W}, {\boldsymbol{\Sigma}},\rho)  ) + log(p(\boldsymbol{W} | {\boldsymbol{\Sigma}}, \vomega^{2})) + log(p(\vomega^{2} |\lambda^2)) + log(p({\boldsymbol{\Sigma}} )) \big] -  E_{q}[log(q(\vtheta))]
	\end{align*}
	\normalsize
Now, taking the expectation with respect to $q$ for each component  in above, we have:
\small
	\begin{align*}
 E_q(log(p(\boldsymbol{Y} | \boldsymbol{W}, {\boldsymbol{\Sigma}},\rho)  ))  & = E_q \bigg[ -\frac{n}{2} log |(D_A-\rho A)^{-1}\otimes{\boldsymbol{\Sigma}}|- \frac{1}{2}\sum_{\ell=1}^n ( \boldsymbol{y}_{\ell} - \boldsymbol{W}^T\boldsymbol{x}_{\ell})^T [(D_A-\rho A)\otimes{\boldsymbol{\Sigma}}^{-1}]\\
 &(\boldsymbol{y}_{\ell} - \boldsymbol{W}^T\boldsymbol{x}_{\ell}) \bigg]\\
 & = -\frac{n}{2} log |(D_A- \rho A)^{-1}\otimes\mu_{q({\boldsymbol{\Sigma}})}| - E_q\bigg[\frac{1}{2}\sum_{\ell=1}^n ( \boldsymbol{y}_{\ell} - \boldsymbol{W}^T\boldsymbol{x}_{\ell})^T [(D_A-\rho A)\\
 &\otimes{\boldsymbol{\Sigma}}^{-1}](\boldsymbol{y}_{\ell} - \boldsymbol{W}^T\boldsymbol{x}_{\ell})\bigg]\\
   E_q(log(p(\boldsymbol{W} | {\boldsymbol{\Sigma}}, \vomega^{2}))) & = E_q \bigg[ \sumid \sum_{j=1}^{\frac{c}{2}} \left({-\frac{1}{2}} log|\omega_{i}^2{\boldsymbol{\Sigma}}|-\frac{1}{2}\tilde{W}_{i,j}^T(\omega_i^2{\boldsymbol{\Sigma}})^{-1}\tilde{W}_{i,j}\right)\bigg] \\
 & = \bigg[ \sumid \sum_{j=1}^{\frac{c}{2}} {-\frac{1}{2}} log|\mu_{q(\omega_{i}^2)} \mu_{q({\boldsymbol{\Sigma}})}|- E_q\left(\frac{1}{2}\tilde{W}_{i,j}^T(\omega_i^2{\boldsymbol{\Sigma}})^{-1}\tilde{W}_{i,j}\right)\bigg]   \end{align*}
 \begin{align*}
 E_q( log(p(\vomega^{2} |\lambda^2))) & = E_q\bigg[ \sumid \Big({\frac{c+1}{2}}log({\lambda^2})  - log(\Gamma(\frac{c+1}{2})) + (\frac{c+1}{2} - 1)log(\omega_i^2)-\frac{\lambda^2}{2}\omega_i^2\Big)\bigg]\\
 &=\bigg[ \sumid \Big({\frac{c+1}{2}} (log({\lambda^2})  - log(2)) - log(\Gamma(\frac{c+1}{2})) + (\frac{c+1}{2} - 1) E_q(log(\omega_i^2))\\
&- \frac{1}{2} \lambda^2 E_q(\omega_i^2)\Big)\bigg] 
  \end{align*}
  \begin{align*}
   E_q(log(p({\boldsymbol{\Sigma}} )) ) &=E_q\bigg[const  - \big({\frac{v+3}{2}} \big)log |{\boldsymbol{\Sigma}}| - \frac{1}{2} tr(S \boldsymbol{\Sigma}^{-1}) \bigg] \\
  &= \bigg[const  - \big({\frac{v+3}{2}} \big)log |\mu_{q({\boldsymbol{\Sigma}})}| - \frac{1}{2} tr(S E_{q(\boldsymbol{\Sigma})}(\boldsymbol{\Sigma}^{-1}))  \bigg]
 \end{align*}
  \normalsize
Now, let's take a look at the $E_q[log(q(\vtheta))]$, which could be written as:
	\small
	\begin{align*}
	E_q[log(q(\vtheta))] & = E_q[log(q(\vW))] + E_q[log(q(\vomega^2)] + E_q[log(q({\boldsymbol{\Sigma}}))] \\  
	& = \sum_{i= 1}^{d}E_q[log(q(vec(\vW_{(i)}^{T})))]  + \sum_{i=1}^{d}E_q[log(q(\omega_i^2)]+E_q[log(q({\boldsymbol{\Sigma}}))] 
	\end{align*}
\normalsize
Where these expectations are evaluated from the forms derived above. Then the lower bound can be written as :
	\small
	\begin{align*}
	\mathcal{L}(q) & = E_{q} \big[ log(p(\boldsymbol{Y} | \boldsymbol{W}, {\boldsymbol{\Sigma}},\rho)  ) + log(p(\boldsymbol{W} | {\boldsymbol{\Sigma}}, \vomega^{2})) + log(p(\vomega^{2} |\lambda^2)) + log(p({\boldsymbol{\Sigma}} )) \big] -  E_{q}[log(q(\vtheta))] \\
	& =  E_{q} \Big[ log(p(\boldsymbol{Y} | \boldsymbol{W}, {\boldsymbol{\Sigma}},\rho)  ) + log(p(\boldsymbol{W} | {\boldsymbol{\Sigma}}, \vomega^{2})) + log(p(\vomega^{2} |\lambda^2)) + log(p({\boldsymbol{\Sigma}} )) \Big] \\
	& - \Big[ \sum_{i= 1}^{d}E_q[log(q(vec(\vW_{(i)}^{T})))]  + \sum_{i=1}^{d}E_q[log(q(\omega_i^2)]+E_q[log(q({\boldsymbol{\Sigma}}))] \Big] \\
	& =  -\frac{n}{2} log \big|(D_A-\rho A)^{-1}\otimes ( S_{q({\boldsymbol{\Sigma}})} / (v_{q({\boldsymbol{\Sigma}})} - 3)\big| - \frac{1}{2}\sum_{\ell=1}^n tr \Big(( \boldsymbol{y}_{\ell} - \boldsymbol{\mu_{q(\boldsymbol{W})}}\boldsymbol{x}_{\ell})^T (\boldsymbol{y}_{\ell} - \boldsymbol{\mu_{q(\boldsymbol{W})}}^T\boldsymbol{x}_{\ell})^T [(D_A-\rho A)\otimes(v_{q({\boldsymbol{\Sigma}})} * S_{q({\boldsymbol{\Sigma}})}^{-1} )]\Big)\\
	& +   \sumid \sum_{j=1}^{\frac{c}{2}}\bigg[ {-\frac{1}{2}} log\Big|\big(\frac{1}{\mu_{q(\eta_i)}} + \frac{1}{\lambda^2 }\big) (S_{q({\boldsymbol{\Sigma}})} / (v_{q({\boldsymbol{\Sigma}})} - 3)\Big|- E_q\left(\frac{1}{2}\tilde{\mu_{q(\boldsymbol{W})}}\tilde{\mu_{q(\boldsymbol{W})}}_{i,j}^T \mu_{q(\eta_i)}(v_{q({\boldsymbol{\Sigma}})} * S_{q({\boldsymbol{\Sigma}})}^{-1} )\right)\bigg]  \\
	&+ \sumid \bigg[ \Big({\frac{c+1}{2}} (log({\lambda^2})  - log(2)) - log(\Gamma(\frac{c+1}{2})) + (\frac{c+1}{2} - 1) (log(\mu_{q(\omega_i^2)}) - \frac{1}{2\mu_{q_{\omega_i^2}}}Var_q[\omega_i^2])- \frac{1}{2}\lambda^2\mu_{q(\omega_i^2) }\Big)\bigg]  \\
	&  -\bigg[\big({\frac{v+3}{2}} \big)log|( S_{q({\boldsymbol{\Sigma}})} / (v_{q({\boldsymbol{\Sigma}})} - 3))| +\frac{1}{2} tr(S (v_{q({\boldsymbol{\Sigma}})} * S_{q({\boldsymbol{\Sigma}})}^{-1} ))  \bigg] \\
	&-  \Bigg[ \sum_{i= 1}^{d})\Big(-\frac{1}{2} log|2 \pi {\boldsymbol{\Sigma}}_{q{(\boldsymbol{W}_{(i)})}}| - \frac{1}{ 2}tr(\boldsymbol{{\boldsymbol{\Sigma}}}_{q{(\boldsymbol{W}_{(i)})}} ^{-1}\boldsymbol{{\boldsymbol{\Sigma}}}_{q{(\boldsymbol{W}_{(i)})}} ) \Big) \\
	  &+   \frac{1}{2} \left( log({\lambda^2} ) -  log(2\pi) \right)  - log(\mu_{q(\omega_i^2})) - \frac{1}{2\mu_{\omega_i^2}}Var_q[\omega_i^2] {\lambda^2}  \left[\frac{1}{2\mu_{q_{(\eta_i)}}^2}  \mu_{q(\eta_i)} - \frac{1}{\mu_{q_{(\eta_i})}^2}  + \frac{\mu_{q(\omega_i^2)}}{2}\right]\\
	   & \frac{v_{q({\boldsymbol{\Sigma}})}}{2} log|S_{q({\boldsymbol{\Sigma}})}| - log(2^{v_{q({\boldsymbol{\Sigma}})}}) - log\Gamma _2(\frac{v_{q({\boldsymbol{\Sigma}})}}{2}) -  \left[ \frac{v+3}{2} log\big(|( S_{q({\boldsymbol{\Sigma}})} / (v_{q({\boldsymbol{\Sigma}})} - 3)|\big)   - \frac{1}{2} tr(S_{q({\boldsymbol{\Sigma}})}(v_{q({\boldsymbol{\Sigma}})} * S_{q({\boldsymbol{\Sigma}})}^{-1} )) \right] \Bigg]  \\
	\end{align*}

\normalsize
\pagebreak

\section*{Web Appendix C: Derivation of the Moment Estimator for $\lambda^{2}$}
	
We have developed a simple approach for tuning the algorithm that is based on a simple moment estimator of $\lambda^{2}$ and does not require multiple runs of the algorithm over different values of $\lambda^{2}$. It is thus well suited to computation on a single processor. Beginning with the model for $\mathbf{W}$ we have
\begin{equation*}
\tilde{W}_{i,j^{*}}| \omega_{i}^{2}, {\boldsymbol{\Sigma}} \stackrel{ind}{\sim}\text{BVN}(\boldsymbol{0},\omega_{i}^2{\boldsymbol{\Sigma}}), 
\end{equation*}
\begin{equation*}
\omega_{i}^{2}|\lambda^{2} \stackrel{iid}{\sim} \text{Gamma}
(\frac{c+1}{2},\lambda^{2}/2), \,\, {\boldsymbol{\Sigma}} \sim \text{Inv-Wishart}(v,\boldsymbol{S})
\end{equation*}

From this we have 
$$
E[W_{i,j}^{2}] = E[E[W_{i,j}^{2}|\omega_{i}^2, {\boldsymbol{\Sigma}}]] = E[VAR[W_{i,j}|\omega_{i}^2, {\boldsymbol{\Sigma}}] +  E^{2}[W_{i,j}|\omega_{i}^2, {\boldsymbol{\Sigma}}]] 
$$
$$
= E[VAR[W_{i,j}|\omega_{i}^2, {\boldsymbol{\Sigma}}] ] 
=
\begin{cases}
E[\omega_{i}^2{\boldsymbol{\Sigma}}_{11}], & \text{if coefficient is for the left hemisphere}\\
E[\omega_{i}^2{\boldsymbol{\Sigma}}_{22}], & \text{if coefficient is for the right hemisphere}
\end{cases}
$$
with $\boldsymbol{S} = \boldsymbol{I}$ we have $E[W_{i,j}^{2}] = \frac{c+1}{\lambda^{2}} \frac{1}{\nu - 3}$. We then use the ridge estimator $\mathbf{\hat{W}}_{R}$ obtained to initialize the VB algorithm to setup a moment equation 
$$
\frac{c+1}{\lambda^{2}} \frac{1}{\nu - 3} = \frac{1}{dc}\sum_{i,j}\hat{\mathbf{W}}_{R_{i,j}}^{2}
$$
and solving yields
$$
\lambda^{2} = \frac{dc(c+1)}{\nu-3}  \left(\sum_{i,j}\hat{\mathbf{W}}_{R_{i,j}}^{2}\right)^{-1}.
$$	
We modify this equation so that it can apply more generally by replacing $\nu -3$ with $\max\{1, \nu-3\}$
which yields
$$
\lambda^{2} = \frac{dc(c+1)}{\max\{1, \nu-3\}}  \left(\sum_{i,j}\hat{\mathbf{W}}_{R_{i,j}}^{2}\right)^{-1}.
$$

\pagebreak
\section*{Web Appendix D: Simulation Study Examining Empirical FDR}
 
\captionsetup{width=16cm}
\begin{center}
\begin{longtable}{l c c}
\caption{ {Empirical proportion of false discoveries in 100 simulation replicates when the expected Bayesian FDR is controlled at $\alpha = 0.05$ leading to the threshold for posterior probabilities $\phi_{\alpha=0.05}$. Note that the values of $c^{*}$ are obtained after transforming the response matrix $\mathbf{Y}$ so that its columns are centered and scaled.} }
\label{Sim1_Results} \\
\hline
&{\centering MCMC} &{\centering VB} \\
\hline  
$c^{*}$ &Empirical FDR &Empirical FDR \\
\hline
\endfirsthead
\multicolumn{3}{l}%
{\tablename\ \thetable\ -- \textit{Continued from previous page}} \\
\hline
&{\centering MCMC} &{\centering VB} \\
\hline
$c^{*}$ &Empirical FDR &Empirical FDR \\
\hline
\hline
\endhead
\hline \multicolumn{3}{r}{\textit{Continued on next page}} \\
\endfoot
\hline
\endlastfoot
0.001 &1.00 &1.00 \\
0.002 &1.00 &1.00 \\
0.003 &1.00 &1.00 \\
0.004 &1.00 &0.99 \\
0.005 & 0.94& 0.81\\
0.006 & 0.56&0.56 \\
0.007 & 0.20&0.37 \\
0.008 & \,\,0.058&0.25 \\
0.009 &\,\,0.019&0.17 \\
0.010 &\,\,\, 0.0067&0.12 \\
0.020 & 0.00&\,\,\, 0.0065 \\
0.030 & 0.00&\,\,\, 0.0012 \\
0.040 & 0.00&0.00 \\
0.050 & 0.00&0.00 \\
0.060 & 0.00&0.00\\
0.070 & 0.00&0.00\\
0.080 & 0.00&0.00 \\
\end{longtable}
\end{center}
\normalsize

\pagebreak
\section*{Web Appendix E: MCMC Trace Plots and Convergence Diagnostics}

\begin{figure}[h]
\centering
\caption{Left Panels displays MCMC trace plots representing  5 randomly selected elements of $\bf{W}$. The right panels display the evolution of Gelman and Rubin shrink factor as the number of iterations increase and GB is the point estimate of Gelman and Rubin's convergence diagnostic statistic.}
\includegraphics[scale=0.6]{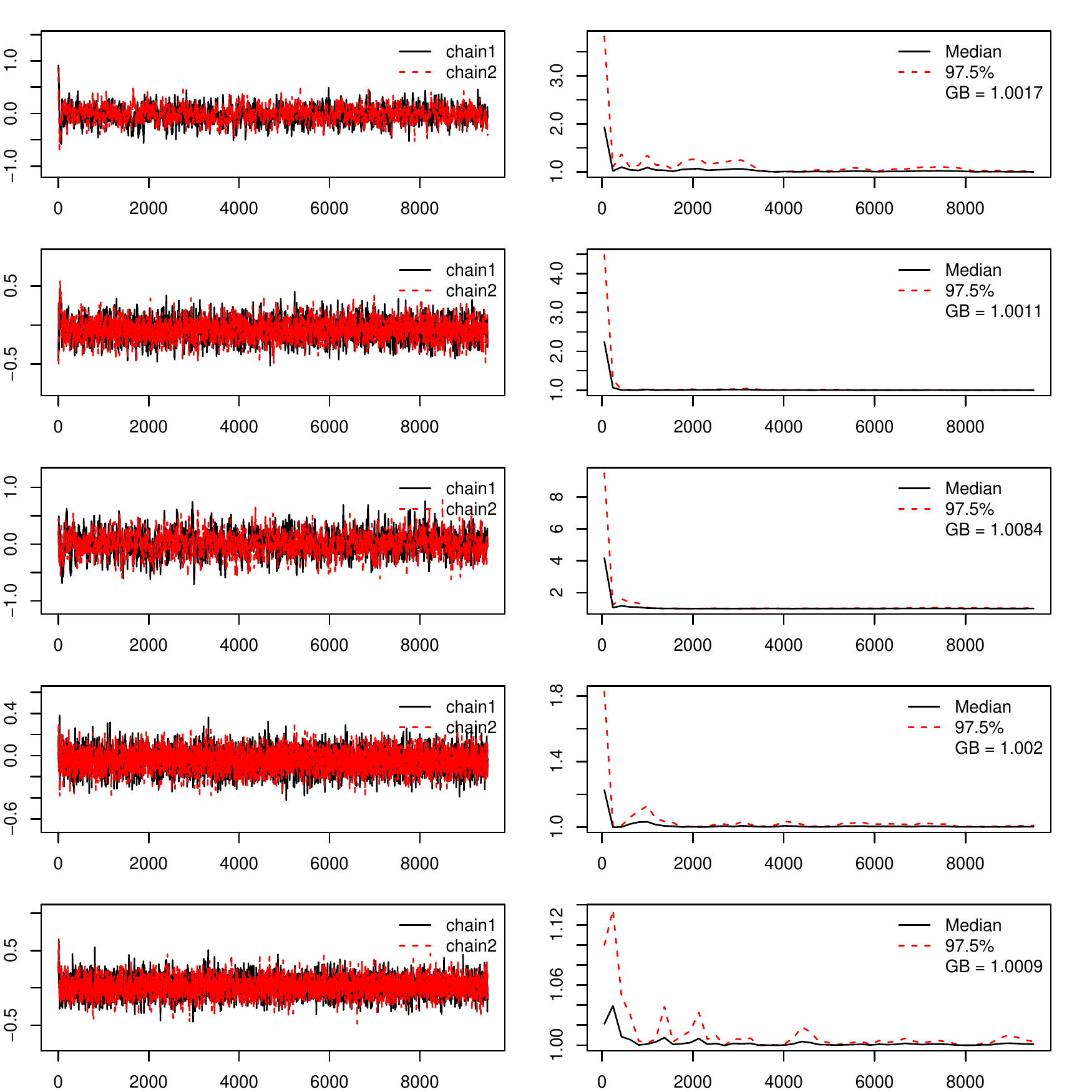}
\label{waicf}
\end{figure}

\begin{figure}[h]
\centering
\caption{Left Panels displays MCMC trace plots representing  ${\boldsymbol{\Sigma}}$. The right panels display the evolution of Gelman and Rubin's shrink factor as the number of iterations increase and GB is the point estimate of Gelman and Rubin's convergence diagnostic statistic.}
\includegraphics[scale=0.6]{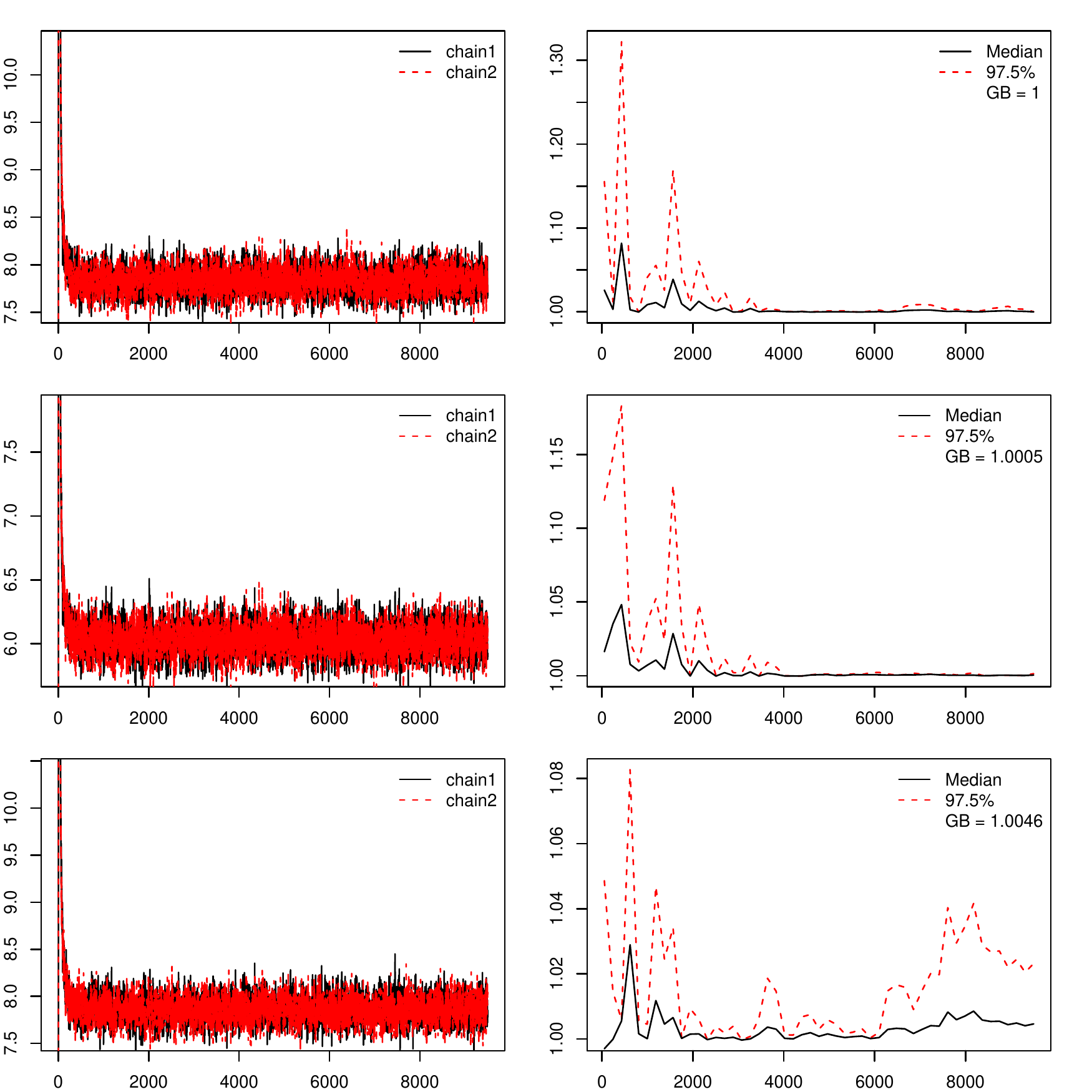}
\label{waicf}
\end{figure}
	
	\pagebreak
\section*{Web Appendix F: Posterior Summaries for APOE SNP rs405509}

\captionsetup{width=16cm}
\small
\begin{center}
\begin{longtable}{p{40mm}llllll}
\caption{ {ADNI-1 Study: posterior means and $95\%$ equal-tail credible intervals for a subset of the ROIs and their association with APOE SNP rs405509.} }
\label{APOECI} \\
\hline
  & \multicolumn{2}{l}{ \textbf{Spatial(MCMC)}}
  & \multicolumn{2}{l}{ \textbf{Spatial(MFVB)}}
 &\multicolumn{2}{l}{ \textbf{Non-Spatial(MCMC)}}\\
  
 Region & Mean & $95\%$ CI & Mean & $95\%$ CI & Mean & $95\%$ CI\\ 
\hline
\endfirsthead
\multicolumn{7}{l}%
{\tablename\ \thetable\ -- \textit{Continued from previous page}} \\
\hline
   & \multicolumn{2}{l}{ \textbf{Spatial(MCMC)}}
  & \multicolumn{2}{l}{ \textbf{Spatial(MFVB)}}
 &\multicolumn{2}{l}{ \textbf{Non-Spatial(MCMC)}}\\

 Region & Mean & $95\%$ CI  & Mean & $95\%$ CI & Mean & $95\%$ CI \\ 
\hline
\hline
\endhead
\hline \multicolumn{5}{l}{\textit{Continued on next page}} \\
\endfoot
\hline
\endlastfoot

 Amygdala volume (L) & 0.09 & [-0.03,0.22] &  0.17& [0.08,0.28]& 0.12 & [0.02,0.23] \\ 
  Amygdala volume (R) & 0.03 & [-0.09,0.16] & 0.10& [0.02,0.21] &0.07 & [-0.03,0.17] \\ 
   Cerebral cortex volume (L) & 0.03 & [-0.08,0.13] & 0.10& [0.02,0.21]& 0.06 & [-0.04,0.16]\\ 
  Cerebral cortex volume (R) & 0.03 & [-0.07,0.14]  &0.13 &[0.03,0.23] & 0.06 & [-0.03,0.16] \\ 
   Cerebral white matter volume (L) & 0.09 & [-0.04,0.22] &  0.19 & [0.10,0.28]& 0.13 & [0.03,0.23]\\ 
  Cerebral white matter volume (R) & 0.02 & [-0.11,0.15] & 0.09 & [0.00,0.18]& 0.05 & [-0.04,0.16]\\ 
  Entorhinal cortex thickness (L) & 0.03 & [-0.08,0.15] & 0.10 &[0.01,0.19]& 0.06 & [-0.04,0.16] \\ 
  Entorhinal cortex thickness (R) & 0.05 & [-0.07,0.17] & 0.10&  [0.00,0.18]& 0.08 & [-0.02,0.18]\\ 
   Fusiform gyrus thickness (L) & 0.11 & [-0.02,0.23] & 0.19& [0.09,0.28]& 0.13 & [0.03,0.23] \\ 
  Fusiform gyrus thickness (R) & 0.07 & [-0.06,0.19] &  0.17 & [0.06,0.27]& 0.10 & [-0.01,0.20] \\ 
  Hippocampus volume (L) & 0.12 & [-0.02,0.26] &0.22 &[0.14,0.34]& 0.15 & [0.05,0.26] \\ 
  Hippocampus volume (R) & 0.07 & [-0.08,0.21] &0.16&[0.06,0.26]& 0.10 & [0.00,0.20]  \\ 
   Inferior lateral ventricle volume (L) & -0.13 & [-0.24,-0.01] &-0.14&[-0.24,-0.05]& -0.08 & [-0.18,0.02]\\
  Inferior lateral ventricle volume (R) & -0.08 & [-0.20,0.04] &-0.08&[-0.17,0.01]& -0.03 & [-0.14,0.07] \\
  Inferior parietal gyrus thickness (L) & 0.08 & [-0.02,0.19] &0.15&[0.05,0.24]& 0.10 & [0.00,0.21] \\ 
  Inferior parietal gyrus thickness (R) & 0.11 & [0.01,0.21] &0.20&[0.11,0.29]& 0.13 & [0.03,0.23] \\ 
   Inferior temporal gyrus thickness (L) & 0.10 & [0.00,0.21] &0.18&[0.09,0.26]& 0.12 & [0.02,0.22] \\ 
  Inferior temporal gyrus thickness (R) & 0.07 & [-0.04,0.17] &0.15&[0.06,0.23]& 0.10 & [-0.01,0.19] \\ 
  Lateral ventricle volume (L) & -0.07 & [-0.17,0.03] &-0.05&[-0.13,0.02]& -0.02 & [-0.12,0.08] \\ 
  Lateral ventricle volume (R) & -0.04 & [-0.14,0.06] &-0.01&[-0.09,0.08]& -0.00 & [-0.10,0.10] \\ 
  Caudal anterior cingulate, isthmus cingulate, posterior cingulate, and rostral anterior cingulate mean thickness (L) & 0.04 & [-0.06,0.14] &0.09&[0.00,0.18]& 0.07 & [-0.03,0.17] \\ 
  Caudal anterior cingulate, isthmus cingulate, posterior cingulate, and rostral anterior cingulate mean thickness (R) & 0.01 & [-0.09,0.11] &0.06&[-0.02,0.15]& 0.05 & [-0.05,0.15]\\ 
  
  Caudal midfrontal, rostral midfrontal, superior frontal, lateral orbitofrontal, and medial orbitofrontal gyri and frontal pole mean thickness (L) & 0.10 & [-0.02,0.22] &0.18&[0.10,0.27]& 0.13 & [0.03,0.23] \\ 
  
   Caudal midfrontal, rostral midfrontal, superior frontal, lateral orbitofrontal, and medial orbitofrontal gyri and frontal pole mean thickness (R) & 0.12 & [0.00,0.25] &0.23&[0.15,0.32]& 0.15 & [0.05,0.26] \\ 
   
   Inferior temporal, middle temporal, and superior temporal gyri mean thickness (L) & 0.10 & [-0.01,0.21] &0.17&[0.08,0.26]& 0.12 & [0.02,0.22] \\ 
   
   Inferior temporal, middle temporal, and superior temporal gyri mean thickness (R) & 0.10 & [-0.01,0.21] &0.20&[0.11,0.30]& 0.13 & [0.02,0.22] \\ 
   
  Fusiform, parahippocampal, and lingual gyri, temporal pole and transverse temporal pole mean thickness (L) & 0.12 & [0.01,0.24] &0.22&[0.13,0.32]& 0.14 & [0.04,0.24]  \\ 
  
  Fusiform, parahippocampal, and lingual gyri, temporal pole and transverse temporal pole mean thickness (R) & 0.10 & [-0.01,0.21] &0.20&[0.11,0.29]& 0.12 & [0.02,0.22] \\ 
  
  Inferior and superior parietal gyri, supramarginal gyrus, and precuneus mean thickness (L) & 0.09 & [-0.02,0.21] &0.16&[0.07,0.25]& 0.12 & [0.02,0.22] \\

  Inferior and superior parietal gyri, supramarginal gyrus, and precuneus mean thickness (R) & 0.11 & [0.00,0.22] &0.20&[0.11,0.29]& 0.13 & [0.03,0.23]\\ 
  
   Precentral and postcentral gyri mean thickness (L) & 0.10 & [-0.01,0.20] &0.16&[0.08,0.25]& 0.12 & [0.02,0.22]  \\
   
  Precentral and postcentral gyri mean thickness (R) & 0.09 & [-0.02,0.20] &0.17&[0.09,0.26]& 0.12 & [0.01,0.22] \\
  
  Inferior temporal, middle temporal, superior temporal, fusiform, parahippocampal, and lingual gyri, temporal pole and transverse temporal pole mean thickness (L) & 0.11 & [0.01,0.22] &0.20&[0.11,0.28]& 0.13 & [0.03,0.23] \\

  Inferior temporal, middle temporal, superior temporal, fusiform, parahippocampal, and lingual gyri, temporal pole and transverse temporal pole mean thickness (R) & 0.10 & [0.00,0.21] &0.21&[0.12,0.29] &0.12 & [0.02,0.23] \\ 
  
  Middle temporal gyrus thickness (L) & 0.08 & [-0.03,0.19] &0.15&[0.06,0.23]& 0.11 & [0.01,0.21] \\
  
  Middle temporal gyrus thickness (R) & 0.09 & [-0.01,0.20] &0.18&[0.11,0.27]& 0.12 & [0.02,0.22] \\
  
  Parahippocampal gyrus thickness (L) & 0.01 & [-0.10,0.11] &0.09&[-0.00,0.21]& 0.03 & [-0.07,0.14] \\
  
  Parahippocampal gyrus thickness (R) & 0.07 & [-0.03,0.18] &0.18&[0.08,0.29]& 0.09 & [-0.01,0.19] \\ 
  
   Postcentral gyrus thickness (L) & 0.13 & [0.03,0.24] &0.23&[0.13,0.33]& 0.14 & [0.03,0.23]  \\ 
   
  Postcentral gyrus thickness (R) & 0.09 & [-0.01,0.20] &0.18&[0.08,0.28]& 0.11 & [0.01,0.21] \\
  
  Posterior cingulate thickness (L) & 0.05 & [-0.07,0.16] &0.11&[0.02,0.20]& 0.07 & [-0.03,0.17] \\
  
  Posterior cingulate thickness (R) & 0.07 & [-0.04,0.18] &0.15 &[0.05,0.24]& 0.10 & [0.00,0.20] \\

   Precentral gyrus thickness (L) & 0.06 & [-0.05,0.17] &0.10&[0.02,0.21]& 0.09 & [-0.01,0.19] \\ 
   
  Precentral gyrus thickness (R) & 0.08 & [-0.03,0.19] &0.15&[0.06,0.25]& 0.11 & [0.01,0.21] \\ 
  
  Precuneus thickness (L) & 0.08 & [-0.02,0.19] &0.14&[0.06,0.23]& 0.10 & [0.00,0.20]  \\ 
  
  Precuneus thickness (R) & 0.09 & [-0.01,0.19] &0.17&[0.08,0.25]& 0.11 & [0.01,0.21] \\ 
  
  Superior frontal gyrus thickness (L) & 0.10 & [0.00,0.20] &0.16&[0.08,0.25]& 0.12 & [0.02,0.22] \\
  
  Superior frontal gyrus thickness (R) & 0.12 & [0.02,0.22] &0.19&[0.11,0.28]& 0.14 & [0.04,0.24] \\
  
  Superior parietal gyrus thickness (L) & 0.07 & [-0.02,0.18] &0.13&[0.05,0.21]& 0.10 & [0.00,0.20]\\
  
  Superior parietal gyrus thickness (R) & 0.08 & [-0.02,0.18] &0.14&[0.06,0.22]& 0.11 & [0.01,0.21]\\ 
  
  Supramarginal gyrus thickness (L) & 0.13 & [0.02,0.23] &0.21&[0.13,0.29]& 0.14 & [0.04,0.25] \\ 
  
  Supramarginal gyrus thickness (R) & 0.14 & [0.04,0.24] &0.24&[0.17,0.33]& 0.15 & [0.05,0.25] \\ 
  
   Superior temporal gyrus thickness (L) & 0.07 & [-0.03,0.18] &0.14&[0.06,0.22]& 0.10 & [0.00,0.20]\\ 
   
  Superior temporal gyrus thickness (R) & 0.10 & [-0.01,0.22] &0.22&[0.14,0.31]& 0.13 & [0.03,0.23] \\ 
  
  Temporal Pole thickness (L) & 0.03 & [-0.07,0.13] &0.09&[0.02,0.18]& 0.06 & [-0.04,0.16] \\
  
  Temporal Pole thickness (R) & -0.02 & [-0.11,0.08]&0.06&[-0.01,0.15] & 0.02 & [-0.08,0.12]\\ 
\end{longtable}
\end{center}
\normalsize

\pagebreak
\section*{Web Appendix G: The WAIC for the Spatial Model Applied to the ADNI-1 Data.}

\begin{figure}[htp]
\centering
\caption{ADNI-1 Data - the WAIC for the spatial model implemented with MCMC for various values of $\lambda^{2}$ and $\rho$.}	
\includegraphics[scale=0.6]{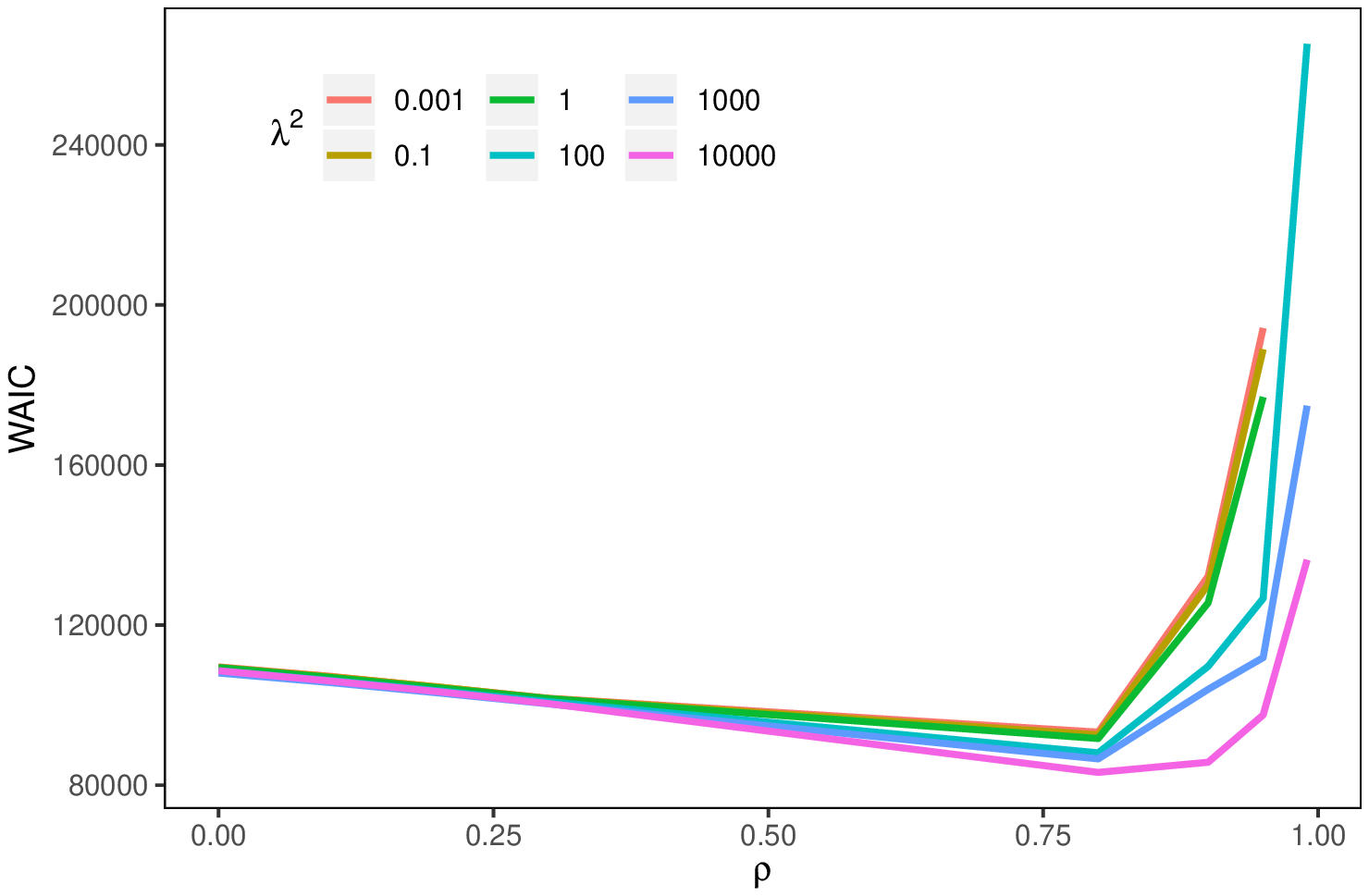}
\label{snpregion}
\end{figure}

\pagebreak
\section*{Web Appendix H: Regularization Paths - Supramarginal Gyrus (Left) and Superior Temporal Gyrus (Left) }

\begin{figure}[htbp]
\centering
\begin{tabular}{cc}
\includegraphics[scale=0.45]{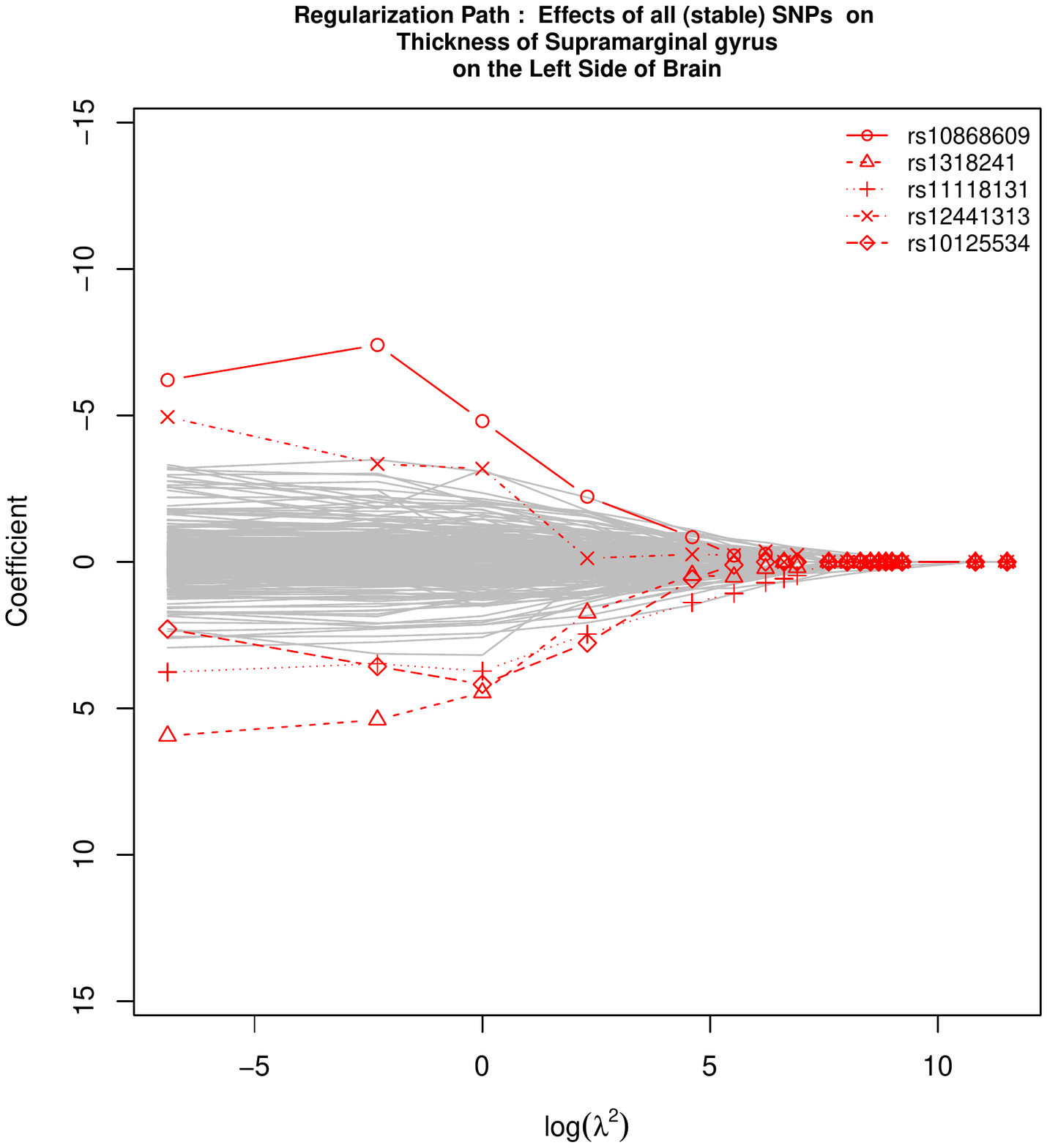} &
\hspace{-2.75em} 
\includegraphics[scale=0.45]{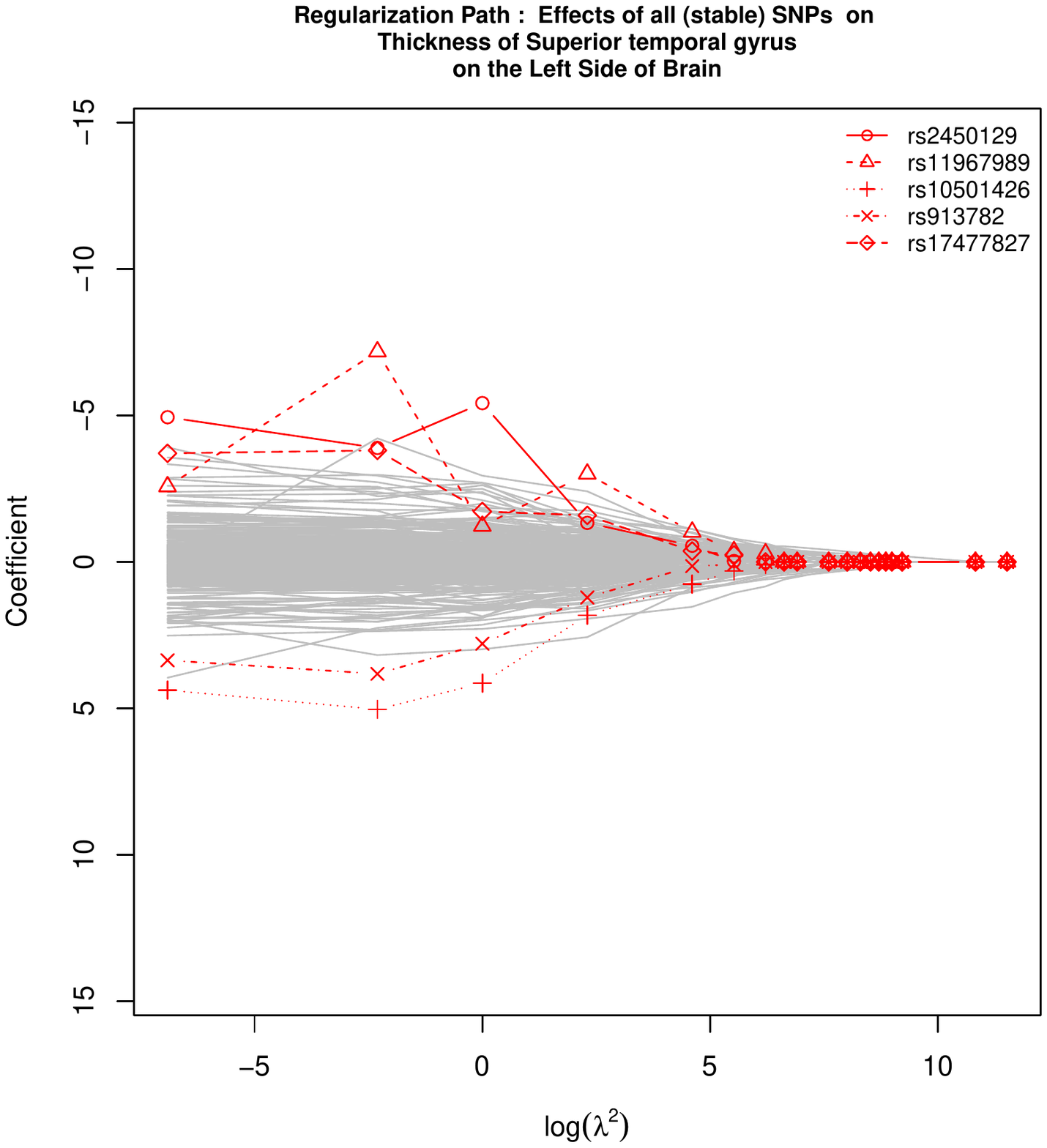}  
\end{tabular}
\caption{ADNI-1 Data: regularization paths showing the posterior mean estimates for varying $\lambda^{2}$ for all SNPs for the thickness of the supramarginal gyrus on the left side of the brain and the thickness of the superior temporal gyrus on the left side of the brain.}
\end{figure}

\end{document}